\documentclass{emulateapj}

  \usepackage{times}
  \usepackage{rotating}
  \usepackage{graphicx}
  \usepackage{dcolumn}
  \usepackage{natbib}

  \bibpunct{(}{)}{;}{a}{}{,} 

  \newlength{\colwidth}
  \newlength{\fullwidth}
  \newcommand{\combo}{{\sc COMBO-17}}
  \newcommand{\gems}{{\sc GEMS}}
  \newcommand{\goods}{{\sc GOODS}}
  
  \newcommand{\re}{$r_{1/2}$}
  \newcommand{\mue}{$\mu_{1/2}$}
  \newcommand{\V}{\mbox{F606W}}
  \newcommand{\Z}{\mbox{F850LP}}

\shorttitle{Colors of AGN host galaxies at $0.5<z<1.1$ from GEMS}
\shortauthors{S.F.S\'anchez et al.}

\begin{document}

  \title{Colors of AGN host galaxies at $0.5<z<1.1$ from GEMS}

  \author{ S.~F.~S\'anchez\altaffilmark{1}, K.~Jahnke\altaffilmark{1},
    L.~Wisotzki\altaffilmark{1,2}, D.~H.~McIntosh\altaffilmark{3}, E.~F.~Bell\altaffilmark{4},
     M.~Barden\altaffilmark{4},
    S.~V.~W.~Beckwith\altaffilmark{5}, A.~Borch\altaffilmark{4},
    J.~A.~R.~Caldwell\altaffilmark{6} , B.~H\"au{\ss}ler\altaffilmark{4},
    S.~Jogee\altaffilmark{6} , K.~Meisenheimer\altaffilmark{4},
    C.~Y.~Peng\altaffilmark{7}, H.-W.~Rix\altaffilmark{4} ,
    R.~S.~Somerville\altaffilmark{6}, C.~Wolf\altaffilmark{8}}

  \altaffiltext{1}{Astrophysikalisches Institut Potsdam, An der Sternwarte 16, 14482 Potsdam, Germany}
  \altaffiltext{2}{Universit\"at Potsdam, Am Neuen Palais 10, 14469 Potsdam, Germany}
  \altaffiltext{3}{Department of Astronomy, University of Massachusetts, 710 North Pleasant Street, Amherst, MA 01003, USA}
  \altaffiltext{4}{Max-Planck-Institut f\"ur Astronomie, K\"onigstuhl 17, 69117 Heidelberg, Germany}
  \altaffiltext{5}{Space Telescope Science Institute \& Johns Hopkins
  University, 3700 San Martin Drive, Baltimore MD, 21218, USA}
  \altaffiltext{6}{Space Telescope Science Institute, 3700 San Martin Drive,
  Baltimore MD, 21218, USA}
  \altaffiltext{7}{Steward Observatory, University of Arizona, 933 N.\ Cherry Ave., Tucson AZ, 85721, USA}
  \altaffiltext{8}{Department of Physics, Denys Wilkinson Bldg., University of Oxford, Keble Road, Oxford, OX1 3RH, UK}

  \email{ssanchez@aip.de}

  \date{Received / Accepted}


  \begin{abstract}
    We present the results from a study of the host galaxies of 15 optically
    selected AGNs with $0.5<z<1.1$ from \gems. \gems\ is a Hubble Space
    Telescope imaging survey of a $\sim 28\arcmin\times 28\arcmin$ contiguous
    field centered on the Chandra Deep Field South in the F606W and F850LP
    filter bands.  It incorporates the SEDs and redshifts of $\sim$10~000
    objects, obtained by the \combo~ project.  We have detected the host
    galaxies of all 15 AGNs in the F850LP-band (and 13/15 in the F606W-band),
    recovering their fluxes, morphologies and structural parameters. We find
    that 80~\% of the host galaxies have early-type (bulge-dominated)
    morphologies, while the rest have structures characteristic of late-type
    (disk-dominated) galaxies.  We find that 25~\% of the early types, and
    30~\% of the late types, exhibit disturbances consistent with galaxy
    interactions.  The hosts show a wide range of colors, from those of
    red-sequence galaxies to blue colors consistent with ongoing star
    formation.  Roughly 70~\% of the morphologically early-type hosts have
    rest-frame blue colors, a much larger fraction than those typical of
    non-active morphologically early-type galaxies in this redshift and
    luminosity range.  Yet, we find that the early-type hosts are structurally
    similar to red-sequence ellipticals, inasmuch as they follow an absolute
    magnitude versus half-light size correlation that are consistent with the
    mean relation for early-type galaxies at similar redshifts.
  \end{abstract}


  \section{Introduction}\label{sec:intro}

  There is now a wide agreement that most if not all galaxies with bulges
  harbor massive black holes in their centers \citep[e.g.,][]{mago98,korm01}.
  However, only a fraction of the galaxies harbor a quasar-like active nucleus
  (AGN). AGNs are often seen as compact sources in the nucleus of galaxies,
  with strong highly-ionized broad and narrow emission lines. They are thought
  to be powered by a central supermassive black hole, fed by the accretion of
  gas from the inner $\sim$100 pc of the host galaxy \citep{anto93,urry95}.
  While both a massive black hole and a reservoir of gas are believed to be
  important for triggering high level AGN activity, it is as yet unclear
  which conditions are needed to trigger such activity. Fundamental questions
  remain regarding the relationship between the AGN, its host galaxy, and star
  formation, the gas inflow rates needed to fuel the nuclear source, and the
  relevant fueling mechanisms on different physical scales.  The morphology
  and spectral properties of AGN host galaxies over different redshift regimes
  can bring important insights to these issues.

  According to the currently favoured view, galaxy-galaxy interactions or
  mergers are important mechanisms for (re-)igniting AGN
  \citep[e.g.,][]{cana01}, except perhaps for very low-level nuclear activity
  \cite[e.g.,][]{coll03,ho03}. Mergers and interactions can induce large-scale
  radial motion of the interstellar gas within the AGN host, and trigger its
  infall into the inner regions to be accreted at later times by the AGN.
  \cite{sand88} first proposed an evolutionary scenario in which galaxy
  encounters could produce ultraluminous infrared galaxies (ULIRGs), that
  would evolve into AGNs. 
  This hypothesis is supported by several studies which found a large fraction
  of merger/interacting systems in the hosts of high-luminosity AGNs
  \citep[e.g.][]{bahc95b, bahc97, hutc97, cana01, sanc03b}. Yet,  
  \cite{dunl03} have not found merger/interactions in their sample of low-$z$
  AGNs. An important limitation of these previous studies is the lack of a
  well-defined comparison sample of inactive galaxies, whose data were
  obtained and analyzed in similar way to the active ones.
  
  A severe practical problem for the study of the hosts of AGNs is the
  presence of the bright nucleus, which has to be carefully removed. This
  strong point-like source, as bright or even brighter than the galaxy,
  contaminates the host flux in all the areas covered by the point spread
  function (PSF).  It also reduces the SNR (signal-to-noise ratio) within the
  host. This affects strongly the reliability of any morphological
  classification of the host. Consecuently, ground-based studies are normally
  limited to low-redshift ($z<0.5$), where the typical size of the host is
  larger than the seeing disc \citep[e.g.][]{dunl93, tayl96, mcle94a, mcle95a,
    jahn03e}, or to wavelength ranges where the host emission dominates the
  AGN (e.g., near-infrared), increasing the contrast \citep[e.g.][]{dunl93,
    mcle94a, falo01, sanc03b}.  The use of HST imaging has clearly increased
  our understanding of the host properties, reducing the effects of
  contamination from the nucleus due to the very narrow intrinsic PSF
  \citep[e.g.][]{mclu00, paga03}. However, most published HST host studies
  were based on WFPC2 imaging with short exposures due to the high AGN flux,
  with a limited determination/treatment of the PSF.
  
  An additional limitation of previous studies has been the sample selection.
  Several samples have been selected {\it ad hoc}, for testing specific
  hypotheses \citep[e.g.,][]{dunl93, cana01}. Other samples are based on
  unclear selection criteria \citep[e.g.,][]{hutc92, bahc95b} or are clearly
  incomplete and/or statistically small \citep[e.g.,][]{hutc97}. In some
  cases the images were not deep enough for an accurate morphological analysis
  \citep[e.g.,][]{lehn99}. A few studies are based on statistically
  significant, flux-limited samples \citep[e.g.,][]{jahn02a,
    jahn03e}, although the results from these studies are limited by the lack
  of a similarly selected comparison sample of inactive galaxies.

  All these difficulties can be better addressed by the {\it Galaxy Evolution
    from Morphology and SEDs} project \citep[\gems,][]{rix03}, where AGNs have
  been identified by the \combo\ multi-color survey \citep{wolf03a}.  We have
  selected the 15 AGNs with $0.5<z<1.1$ from \gems, coincident with the best
  studied redshift range for the inactive galaxies in that survey. Using an
  accurate determination of the PSF, and a two-dimensional (2D) fitting
  technique, we decouple the host and nuclear components in these objects, and
  compare the properties of the host galaxies with those of inactive galaxies.
  
  The sequence of this article is as follows. In Section 2 and 3 we describe
  the data and the data analysis. In Section 4 we present the results, and we
  discuss these in Section 5. The summary and conclusions are presented in
  Section 6. Throughout this article we assume a $\Lambda$-cosmology, with
  $H_0=70~\mathrm{km~s}^{-1}~\mathrm{Mpc}^{-1}$, $\Omega_{\mathrm{m}}=0.7$ and
  $\Omega_{\Lambda}=0.3$. We have used the AB photometric
  system for the observed magnitudes and colors, and the Vega system for the
  rest-frame colors and absolute magnitudes, unless explicitly stated
  otherwise.

  \begin{table}
  \caption{Properties of the \gems\  low-$z$ AGN Sample}
  \label{tab1}
  \begin{tabular}{cccccc}
  \tableline\tableline
  QSO ID&Tile&$z^1$&$R$-band&F606W-band&F850LP-band\\
  \tableline
34357&  94 &{\it 0.543}&18.73&19.22$\pm$0.06&18.40$\pm$0.06\\ 
41310&  47 &0.548&23.09&23.24$\pm$0.09&22.48$\pm$0.09\\ 
41310&  40 &0.548&23.09&23.20$\pm$0.09&22.44$\pm$0.08\\ 
52963&  50 &0.548&22.69&22.72$\pm$0.09&22.10$\pm$0.08\\ 
36361&  94 &0.549&22.50&23.39$\pm$0.09&22.17$\pm$0.13\\ 
47615&  46 &0.649&22.28&22.08$\pm$0.08&21.36$\pm$0.08\\ 
50415&  85 &{\it 0.664}&22.39&22.75$\pm$0.09&21.70$\pm$0.08\\ 
50415&  51 &{\it 0.664}&22.39&22.74$\pm$0.09&21.66$\pm$0.08\\ 
44126&  42 &0.729&23.03&23.04$\pm$0.09&22.33$\pm$0.09\\ 
42601&  42 &{\it 0.733}&21.02&21.32$\pm$0.08&20.24$\pm$0.07\\ 
48284&  85 &{\it 0.734}&19.05&19.33$\pm$0.06&18.92$\pm$0.06\\ 
39432&  89 &{\it 0.738}&22.02&22.16$\pm$0.08&21.43$\pm$0.08\\ 
31898&  30 &0.812&22.91&23.10$\pm$0.09&21.62$\pm$0.08\\ 
15731&  18 &0.835&20.11&19.77$\pm$0.07&19.43$\pm$0.06\\ 
50997&  85 &{\it 0.837}&20.43&20.60$\pm$0.07&20.38$\pm$0.08\\ 
50997&  51 &{\it 0.837}&20.43&20.43$\pm$0.07&20.22$\pm$0.07\\ 
49298&  52 &{\it 1.031}&19.92&19.89$\pm$0.07&19.47$\pm$0.07\\ 
49298&  85 &{\it 1.031}&19.92&19.90$\pm$0.07&19.56$\pm$0.06\\ 
43151&  95 &{\it 1.037}&22.18&22.25$\pm$0.08&21.78$\pm$0.08\\ 
43151&  45 &{\it 1.037}&22.18&22.38$\pm$0.09&21.92$\pm$0.08\\ 
  \tableline
  \end{tabular}

$^1$ values in italics are spectroscopic redshifts obtained from \cite{szok04}.

  \end{table}

  \section{The data}\label{sec:data}
  
  \gems\ is a two band, F606W and F850LP (similar to the $V$ and $z$-bands),
  HST imaging survey covering 78 pointings using the ACS camera. This survey
  covers a continuous field of $28\arcmin\times 28\arcmin$ in the extended
  Chandra Deep Field South, to a depth of F606W$=$28.3 (5$\sigma$) and
  F850LP$=$27.1 (5$\sigma$) for compact sources. In its central $\sim$1/4,
  \gems\ incorporates ACS imaging from the \goods\ project \citep{giav03}.
  Only the first epoch coverage from \goods\ was used, resulting in images
  that are considerably shallower than the final \goods\ data, but also
  slightly less deep than the surrounding proper \gems\ tiles.  The full
  \gems\ data reduction will be described elsewhere (Caldwell et al., in
  preparation). It comprises the standard data reduction (bias subtraction,
  flat-fielding, flux calibration), drizzling of the data from the original
  0\farcs05/pixel to a final sampling of 0\farcs03/pixel, background
  estimation and variance determination for each pixel.
  As the deblending of nuclear and host galaxy
  components with two-dimensional modelling (the main purpose of this article)
  is sensitive to background errors, we applied an extra procedure to remove
  local background residuals. This includes an iterative masking of all
  objects in the field and the determination of the local background from the
  object-free regions. For each area of $200\times 200$ pixels an average
  from the unmasked pixel was computed, with a subsequent bilinear
  interpolation between these values to yield a background estimate for the
  whole field. The overall properties (pixel size,
  noise pattern, etc.) of the images in both passbands  
  are quite similar, although the F606W images are deeper by 1 mag.

  \begin{figure}
  \includegraphics[angle=-90,width=\hsize]{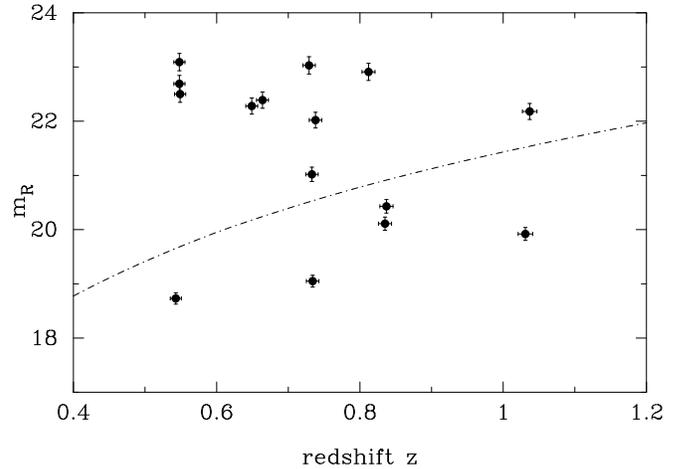}
  \caption{\label{z_R} 
    $R$ band magnitudes of the AGN sample as measured in \combo , plotted
    against \combo\ redshift.  The dot-dashed line corresponds to 
    constant $M_B = -23$ for the mean quasar SED of \cite{elvis94}.
  }
  \end{figure}

  The \gems\ area has been previously studied by \combo, a photometric
  redshift survey based on imaging in 12 medium and 5 broad-band filters
  \citep{wolf03a}.  \gems\ incorporates the redshift and spectral energy
  distribution (SED) classifications of 10~000 objects from \combo, including
  a sample of $\sim$120 AGNs with $0.5<z<5$.  Because of the SED-based
  selection technique within \combo, most of these AGNs are type I, i.e., they
  show broad emission lines.  We introduced a cut at $m_R<24$ Vega mag for SED
  and redshift reliability, defining a subsample of $\sim$80 objects.  For the
  redshift range of our interest ($0.5<z<1.1$), this cut defines a subsample
  of 18 AGNs, and limits the uncertainty of the \combo\ redshift to
  $\Delta z\sim$0.02.  In a final cleaning pass of the sample 
  we excluded, by visual inspection, those objects located at the 
  very edges of the tiles.
  The final sample comprises $\sim$60 AGNs with accurate redshift and good
  imaging, 15 of them at $0.5<z<1.1$. Most of of the finally selected objects
  have absolute magnitudes of around or just below $M_B \simeq -23$, which for
  an Einstein-de Sitter universe with
  $H_0=50~\mathrm{km~s}^{-1}~\mathrm{Mpc}^{-1}$ corresponds to the
  conventional division between high-luminosity QSOs and lower luminosity
  Seyferts, \cite{schm83}.

  This sample allows us, for the first time, to study
  a complete flux-limited sample of AGN hosts, with a
  similarly-selected and studied inactive comparison sample.  
  We have extracted postage-stamp images of $128\times 128$ pixels size 
  ($3\farcs 84\times 3\farcs 84$) of the F606W and F850LP-bands,
  centered on each AGN.
  
  The redshift range has been selected for several reasons. The main
  aim of this selection is that the F606W and F850LP-band sample two
  physically distinct
  spectral ranges. While the F606W-band always samples the wavelength
  range bluewards of the Balmer break ($\sim$4000\AA), and is therefore
  sensitive to OB stars in the galaxies,
  the F850LP-band lies entirely redwards of
  this break, probing the old population of red stars in the galaxies.
  The F606W$-$F850LP colors will therefore allow us to explore the 
  degree of star formation activity in a homogeneus way for the entire sample.
  The angular scales between $z\sim 0.5$ and 1.0 change only by a factor 1.5,
  and cosmological surface brightness dimming is not yet a major effect.  
  Therefore, this selection allows us to perform a structural analysis
  in a homogeneous way to all objects.  Finally, this
  redshift range coincides with the peak of the \gems\ redshift distribution,
  enabling us to access a large and well-defined comparison sample. In a
  separate article we study the host galaxy properties of \gems\ AGN
  at much higher redshifts, $1.8<z<2.75$ \citep{jahn04a}.

  Table \ref{tab1} gives a summary of the properties of the AGNs of our
  sample, including the \combo~ID, the \gems~tile of the analyzed image, the
  redshift and $R$-band photometry from \combo\ and the F606W and F850LP-band
  photometry from the \gems\ data.  Whenever available, we have used the
  spectroscopic redshifts obtained by \cite{szok04} in their optical study of
  the X-ray sources in the Chandra Deep Field South (redshifts in italics in
  table \ref{tab1}). The spectroscopic and \combo\ redshifts agree within the
  expected errors (for a full discussion of redshift accuracy in \combo\ see
  \cite{wolf04}).  Five of the 15 AGNs are located in overlapping tile areas
  and thus received
  double exposure times.
  We have analyzed each individual image and compared the results at the end.
  Figure \ref{z_R} shows the distribution of (\combo\ based) $R$-band Vega 
  magnitudes as a function of redshift.

  \section{Analysis of the data}\label{sec:removal}
  
  There are various problems in the detection and restoration of the hosts of
  AGNs. Most of them relate to the presence of the AGN itself, a strong
  point-like source that, due to the PSF, contaminates the host image. The
  level of contamination depends on the contrast (the flux ratio between the
  nucleus and the extended component), the overall SNR, and the
  accuracy of the PSF determination. Here, we present the method used
  to detect and characterize the AGN hosts, and to estimate the uncertainties.

  \begin{figure}
  \includegraphics[width=\hsize]{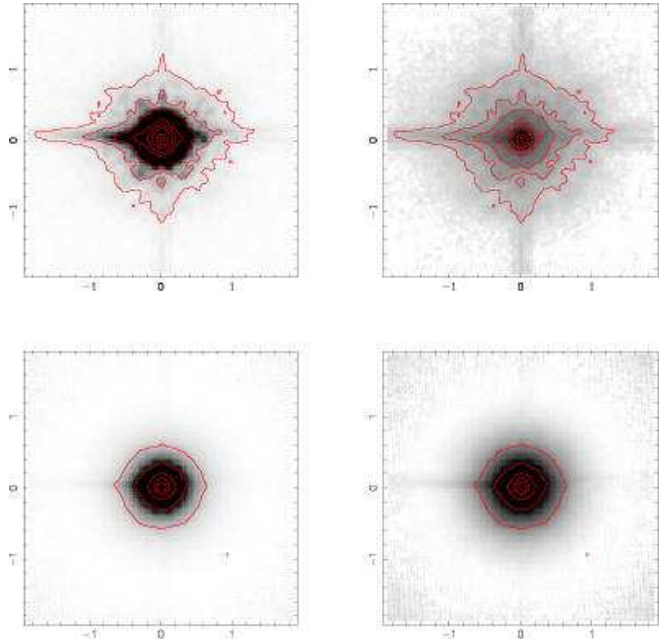}
  \caption{\label{PSF} 
    Average PSF over the whole GEMS field for the F850LP-band, plotted in 
    linear (top-left) and logarithmic scales (top-right).  The variance map of
    the PSF is also plotted, in linear (bottom-left) and logarithmic scales 
    (bottom-right).  The contours are plotted in logarithmic scale, starting
    at 10$^{-5}$ counts for the 1st contour and with a separation of 0.5 dex. 
  }
  \end{figure}

  \subsection{PSF production}\label{sec:psf}
  
  An extensive study of the ACS PSF based on \gems\ data will be presented
  elsewhere 
  (Jahnke et al., in preparation). On the basis of this study, we find that
  the PSF shape varies across the ACS field-of-view, with pixel-to-pixel flux
  variations $\la 20$~\%. This prevents us from using a single mean PSF for our
  analysis. We find that temporal variations are negligible compared to
  spatial variations.  Therefore, we built a local PSF around the position of
  each AGN, averaging the nearest 35 isolated stars within a distance of
  $\sim 40\arcsec$. In order to take the uncertainties of the
  PSF determination into account for later image modelling, we included 
  these uncertainties in the total pixel variance budget. 
  Figure \ref{PSF} shows the mean PSF averaged over the
  whole GEMS field, plotted in both a linear and a logarithmic scale to show
  the spikes and substructures.  We also show the
  variance map of the global mean PSF; each individual PSF has a similar
  associated variance map which was used in the fitting process.

  \subsection{Simple PSF subtraction}\label{sec:psfsub}
  
  A simple method for the detection of the hosts was applied by subtracting
  the PSF scaled to the QSO peak intensity: the PSF was scaled to the central
  flux of the AGN, integrated inside a circular aperture of 4 pixels ($0\farcs
  12$) diameter centered on the nucleus, and then subtracted from the AGN.
  This method is very conservative with respect to the host galaxy detection
  since it implies, by definition, an oversubtraction of the nuclear component
  from the inner regions. It provides us with a non-model-dependent detection
  and a lower limit on the host flux.  In order to quantify the results from
  this detection method we tested it using 200 field stars, covering a
  magnitude range similar to our objects. We find false detections of a host
  contributing more than a 10~\% (5~\%) to the total flux in only 3~\% (12~\%) of
  the field stars.  Here we will assume a \emph{hard} lower limit of 10~\% of
  the total flux for a reliable host detection. We expect no spurious
  detections in our sample of 15 objects with this adopted limit.
  
  Applying this method, we have detected hosts in all the 15 objects for the
  F850LP band, and in 13 of the 15 AGNs for theF606W band images. The two
  objects not detected in the F606W-band, \combo\ 50997 and \combo\ 43151, are
  both at a relatively high redshift. Their hosts are clearly detected in the
  F850LP-band, being most likely compact spheroids (as we will discuss below).
  Fortunately, both objects were observed by \gems\ and \goods, being in an
  overlapping region. In this case we have combined the images obtained from
  both projects, and applied the subtraction analysis. We have found some
  residuals in the images, especially for \combo\ 50997, coincident with the
  F850LP-band residuals by a visual inspection. However, they are still below
  the relibility criteria and we disregard these weak residuals in what
  follows.

  \subsection{2D modeling}\label{sec:galfit}
  
  We performed two-dimensional modelling of the images to recover more accurately the
  nuclear and host fluxes, and to determine some morphological parameters of
  the hosts. The fitting was performed using GALFIT v1.7a \citep{peng02}.
  This software fits galaxy images with a 2D model, which consists of
  a superposition of parametrized fitting functions for point-like and 
  extended sources. A fitting process then optimizes the
  flux scaling factors and morphological parameters such as scale lengths,
  ellipticities and position angles, by means of $\chi^2$ minimization
  operating on the original pixel data.
  
  The AGN images were fitted with a two-component model: a Gaussian function
  with a very narrow width for the nuclear component (a quasi-delta function)
  and a galaxy model for the host. The FWHM of the Gaussian was fixed to the
  average value obained from the fit with a single Gaussian to the stars used
  to create the PSF.  Nearby companions were masked before performing the
  fitting.  We did not preselected a certain galaxy model (disk or spheroid)
  for our objects, since the morphological determination was one of the goals
  of this study.  Instead, we assumed that the hosts could be characterized by
  either of two single models: (1) an exponential disk \citep{free70} for
  disk-dominated galaxies, or (2) a de Vaucouleurs $r^{1/4}$ function
  \citep{deva79} for bulge-dominated galaxies. Due to the complexity of the
  problem, we did not try to decouple the disk from the bulge, which would
  have required a third fitting function into the model. We also performed a
  fit to characterize the host galaxy as a S\`ersic function \citep{sers68},
  which helped in establishing a morphological classification as shown below.

  The 2D fit provided us with a robust method to decouple the nucleus and the
  host, allowing us to obtain the fluxes of both components. 
  For each model, the nucleus image 
  was subtracted from the original image, to provide a complete image of
  the host, including its substructures.  We have obtained the total flux of
  the host by direct measurement of this \emph{restored} image.
  
  Finally, we also performed a surface brightness (SB) profile analysis
  of the images, fixing the ellipticity and the position angle to the values
  obtained from the 2D fitting. The SB profiles were used to visually
  check the results from the 2D modelling. The physical scale and the surface
  brightness of the galaxies were estimated measuring the half-light
  radius (\re) and the SB at this radius (\mue).
  
  Tables \ref{ltable_V} and \ref{ltable_z} summarize the results of this
  analysis.  For the images of each object we list the model of the host
  included in the fitting (disk, de~Vaucouleurs or the S\`ersic index derived
  from the fitting), the derived total, nuclear and host galaxy magnitudes
  from the model, the host-to-nuclear flux ratio (H:N), the axis ratio of the
  host (a/b), the position angle (PA) and the reduced $\chi^2$ of the fit.  We
  also provide the magnitude derived from the peak-scale PSF subtraction
  described above (Sub), the magnitude of the \emph{restored}
  host image (Emp) together with the half-light radius and
  surface brightness.

  \begin{figure}
  \includegraphics[width=\hsize]{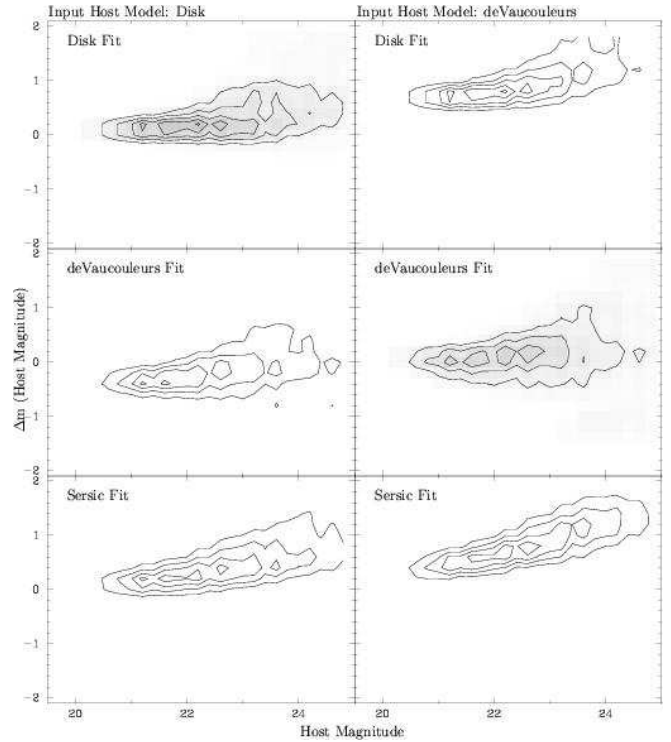}
  \caption{\label{simulations} 
    Density distribution of the differences between the input and the output
    magnitudes of the host galaxies for the two simulated models (disk and
    de Vaucouleurs), when fitted with three different functions: exponential
    disk, de Vaucouleurs and S\`ersic, as described in the text.
    The first contour encircles 90~\% of the objects, with a separation of
    20~\% between each successive contours.
  }
  \end{figure}

  \subsection{Limits of the method: simulations}
  
  To understand the limitations of our methodology we performed simulations.
  We created simulated images that mimic, as much as possible, the real data.
  Due to the large similarities between our F606W and F850LP images, we
  simulated only F850LP images, and translated the results to F606W by a
  simple zero-point shift transformation
  (F606W$_{\mathrm{0}}-$F850LP$_{\mathrm{0}}=$1.6502 mags).  The AGNs were
  simulated with a two component model (nucleus+host), plus noise.  The
  nucleus was simulated by a scaled PSF, using the mean PSF shown in Figure
  \ref{PSF}. The host galaxy was simulated using a single model, either an
  exponential disk or a de~Vaucouleurs function. Clearly this was an
  oversimplification of the true diversity in galaxy morphological types.  
  However, most of our objects can be clearly classified as either bulge- or
  disk-dominated (see below), with very few cases of ambiguous morphological
  type. The simple dichotomy assumed in the simulations is therefore 
  appropriate for the purpose of this paper.

  To reproduce the noise pattern seen
  in the real images, due to the drizzling process on the reduction (Caldwell
  et al., in preparation), we created the noise map in two steps.  A first
  guess of the noise map was built using the noise-free simulated images,
  assuming a Poisson distribution for the noise, and adding the noise due to
  the background and readout noise. Due to the drizzling the noise
  distribution of the real images is narrower than that expected from
  Poissonian statistics, and there are correlations between neighboring
  pixels.  In order to simulate this noise pattern we smoothed the simulated
  noise images by a mean filter using a 3$\times$3 pixels box. Then we
  combined this smoothed noise map with the original Poissonian one, with
  different relative weights, until recovering the observed noise
  distribution. The final noise map consisted of 80\% of the original map plus
  20\% of the smoothed one.
  
  We covered a wide range of parameter space in our simulations, going
  outside the expected boundaries for real data to explore possible
  systematic effects. The simulations range in total magnitudes
  between 18.5 and 25.5, with a range of host-to-nucleus flux ratios
  between 0.1 and 2.0 ($\sim 9$--66~\% of the total flux), and half-light radii
  for the hosts between 0$\farcs$15 and 1$\farcs$0 (between 0.5 and 10 kpc
  at the redshift of our objects). For each set of data we created five
  different realizations, based on different noise maps. We finally
  created a total of 1880 simulated AGN, 940 pairs with similar parameters,
  but different host morphologies. The simulated images were created
  both by the GALFIT and IRAF/ARTDATA packages, without significant
  differences in the final results.
  
  With these simulations we explored the range of parameter space 
  for which our method could robustly reclaim the key parameters 
  (total, nucleus and host fluxes, host scale length, and the contrast
  between host and nucleus). We also searched for possible biases in the 
  derived parameters, and we estimated statistical errors by comparing 
  the input and output
  values. To place limits on our ability to differentiate between
  disk-dominated and bulge-dominated galaxies, we fitted all the simulations,
  like the real data, with three different models for the host galaxy: 
  An exponential disk, a de~Vaucouleurs function, and a S\`ersic function 
  (with a free $n$-index).  The total number of performed fits was 5640.
  A detailed description of these simulations will be presented
  elsewhere (S\'anchez et al, in preparation), here we give only a brief summary. 
  In all cases we
  find that the nuclear magnitude is recovered with higher accuracy 
  than the host magnitude.
  Figure \ref{simulations}
  shows the distribution of the difference between the input and output host
  magnitudes versus the input magnitudes, for the different input and fitted
  models. It is clear from this figure that the best estimation
  of the host magnitude depends on an accurate assessment of the host
  morphological type.
  
  The error in total and nuclear magnitudes is always lower than $\sim$0.12
  magnitudes. On the other hand, the accuracy of host magnitudes 
  depend more on the host flux (SNR) than on other parameters like the
  contrast (host-to-nucleus ratio), or the host physical scale (for
  $r_{1/2}>0\farcs 15$). The flux of the disk hosts is better recovered than
  the flux of the elliptical hosts for the same range of parameters. This is
  due to the ACS PSF following roughly an $r^{1/4}$-profile (Jahnke et al., in
  preparation), which makes it more difficult to disentangle nuclear point sources
  from spheroidal than from disk host. Another reason could be that
  disk galaxies are less concentrated, and therefore, for the same luminosity,
  they have more light in the wings. The error of the host magnitudes ranges
  from $\sim$0.05 mag for the brightest disk hosts 
  (F850LP$_{\mathrm{host}} \sim 20$) to $\sim$0.40 mag for the faintest
  spheroidal hosts (F850LP$_{\rm host} \sim 24$).

  \begin{figure}
  \includegraphics[angle=-90,width=\hsize]{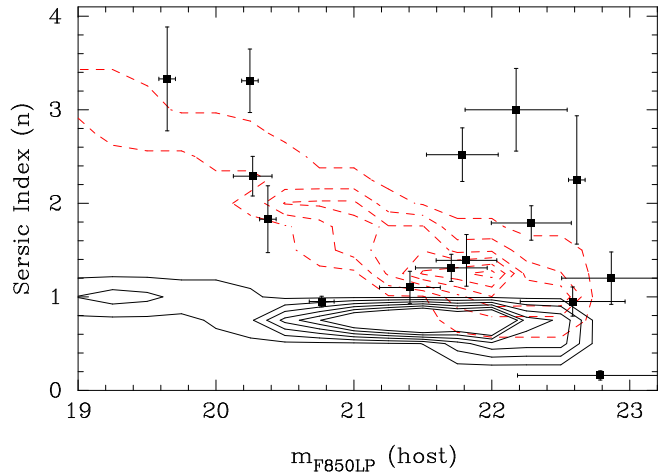}
  \caption{\label{sersic} 
    Density distribution of the output S\`ersic index from 2D fitting to
    simulated disk ($n=1$, solid line) and de Vaucouleurs ($n=4$, dotted line)
    host galaxies. The outer contour encircles 95~\% of the simulations, with
    increments of 15\% between successive contours. The solid squares show
    that distribution for the host galaxies of our AGN sample (values listed
    in table 3). 
  }
  \end{figure}

  It has already been noted \citep{mclu99,sanc03b} that it is not possible to perform
  an accurate morphological classification of host galaxies based simply on
  the goodness of the fit to either disk or de Vaucouleurs models (i.e., a
  comparision of the reduced $\chi^2$). The model fitting is dominated so
  strongly by the nucleus fitting that the derived goodness parameter is
  similar for both models \citep{sanc03b}.  However, we find that we achieve
  a reliable morphological classification using the S\`ersic index from our
  fitting and comparing this to the results of our simulations. In absence of
  the nucleus and without noise, the S\`ersic index should be 1 for a pure disk
  model and 4 for a pure de Vaucouleurs model. The presence of the nucleus and
  the noise alters this result. This classification method was already used by
  \cite{mclu99} and \cite{sanc03b}.
  
  Figure \ref{sersic} shows the distribution of the output S\`ersic indices
  along the simulated host magnitudes for the input disks (solid contours) and
  the input de Vaucouleur spheroids (dashed contours). The S\`ersic indices 
  are close to their input values only for very bright host galaxies and tend
  to become smaller as a consequence of the presence of a central point source.
  Nevertheless, the two classes of disk and de~Vaucouleurs models remain well
  separated unless the host galaxy is very faint. Based on these simulations
  we conclude that we can perform a reliable morphological classification at least
  into disk- or spheroid-dominated types for most of our objects.

\begin{table*}
\footnotesize
\caption{Results from the 2D fitting: F606W-band images}
\label{ltable_V}
\begin{tabular}{ccccccccrccccc}
\tableline\tableline
QSO ID&Tile&Model&Tot&Nuc&Host&H:N&a/b&P.A.&$\chi^2_{\nu}$&Sub&Emp&r$_{\rm 1/2}$&$\mu_{\rm 1/2}$\\ 
\tableline
 34357&94& disk    &19.29&19.51&21.15&0.22&0.87&15.27& 1.57 &21.70&21.96&0.52&23.20\\ 
 &         &deVauc   &19.19&19.53&20.63&0.36&0.92&18.55& 1.57 &21.70&21.82&0.52&23.36\\ 
 &         &1.76&19.27&19.53&20.96&0.26&0.89&16.15& 1.56 &21.70&21.86&0.52&23.27\\ 
 41310&47& disk    &23.32&23.91&24.27&0.71&0.52&44.38& 0.35 &24.20&24.27&0.20&23.47\\ 
 &         &deVauc   &23.24&24.58&23.62&2.42&0.52&45.91& 0.34 &24.20&23.64&0.16&22.68\\ 
 &         &3.77&23.25&24.67&23.60&2.67&0.53&45.88& 0.34 &24.20&23.62&0.16&22.64\\ 
 41310&40& disk    &23.31&23.95&24.19&0.80&0.52&47.72& 0.39 &24.07&24.20&0.16&22.93\\ 
 &         &deVauc   &23.25&24.94&23.51&3.73&0.55&48.67& 0.39 &24.07&23.53&0.11&21.71\\ 
 &         &4.60&23.23&24.96&23.48&3.90&0.53&48.65& 0.39 &24.07&23.50&0.11&21.70\\ 
 52963&50& disk    &22.91&24.19&23.32&2.22&0.43&4.20& 0.55 &23.29&23.34&0.20&22.38\\ 
 &         &deVauc   &22.78&25.05&22.93&7.04&0.38&5.80& 0.53 &23.29&22.97&0.16&21.74\\ 
 &         &1.75&22.81&24.33&23.12&3.04&0.37&5.56& 0.51 &23.29&23.14&0.11&20.91\\ 
 36361&94& disk    &23.39&25.28&23.60&4.69&0.57&$-$9.59& 0.33 &24.91&24.44&0.25&23.96\\ 
{\it mask}&&deVauc   &23.14&26.44&23.17&$>$10&0.59&$-$14.02& 0.34 &24.91&24.10&0.25&24.08\\ 
 &         &0.51&23.42&24.87&23.76&2.77&0.56&$-$7.22& 0.33 &24.91&24.59&0.25&23.92\\ 
 47615&46& disk    &22.11&22.51&23.42&0.43&0.84&$-$80.08& 0.56 &23.57&23.44&0.25&23.06\\ 
 &         &deVauc   &22.07&22.77&22.88&0.90&0.80&$-$78.31& 0.57 &23.57&22.91&0.20&22.57\\ 
 &         &1.59&22.10&22.57&23.26&0.52&0.83&$-$79.25& 0.56 &23.57&23.25&0.20&22.63\\ 
 50415&85& disk    &22.96&23.45&24.08&0.55&0.99&87.91& 0.30 &24.62&24.89&0.38&25.47\\ 
{\it mask}&&deVauc   &22.83&23.53&23.64&0.90&0.93&6.87& 0.31 &24.62&24.60&0.38&25.68\\ 
 &         &0.75&22.96&23.42&24.12&0.52&0.88&$-$57.55& 0.30 &24.62&24.92&0.34&25.15\\ 
 50415&51& disk    &22.84&23.40&23.83&0.67&0.92&72.84& 0.35 &23.74&23.84&0.38&24.44\\ 
{\it mask}&&deVauc   &22.68&23.52&23.36&1.15&0.89&52.31& 0.35 &23.74&23.52&0.34&24.35\\ 
 &         &0.90&22.85&23.39&23.87&0.64&0.91&58.94& 0.34 &23.74&23.88&0.38&24.43\\ 
 44126&42& disk    &23.07&23.16&25.85&0.08&0.56&$-$36.55& 0.42 &25.73&24.96&0.11&23.10\\ 
 &         &deVauc   &23.08&23.24&25.28&0.15&0.20&$-$21.78& 0.42 &25.73&25.37&0.16&23.38\\ 
 &         &0.02&23.05&23.12&26.19&0.05&0.67&$-$65.22& 0.42 &25.73&25.25&0.11&23.55\\ 
 42601&42& disk    &21.35&22.86&21.67&2.99&0.73&23.95& 0.82 &21.69&21.73&0.65&23.43\\ 
 &         &deVauc   &20.83&23.25&20.96&8.24&0.73&25.99& 0.85 &21.69&21.59&0.65&23.62\\ 
 &         &1.69&21.23&22.96&21.48&3.90&0.73&25.22& 0.78 &21.69&21.66&0.65&23.52\\ 
 48284&85& disk    &19.39&19.50&22.03&0.09&0.89&$-$72.63& 2.71 &22.43&21.96&0.38&23.42\\ 
 &         &deVauc   &19.36&19.53&21.48&0.16&0.84&$-$75.80& 2.68 &22.43&22.38&0.29&22.89\\ 
 &         &4.27&19.35&19.52&21.47&0.16&0.94&$-$62.78& 2.68 &22.43&22.45&0.34&23.25\\ 
 39432&89& disk    &22.28&22.46&24.38&0.17&0.92&$-$45.11& 0.33 &24.92&25.18&0.25&24.83\\ 
 &         &deVauc   &22.19&22.44&23.95&0.24&0.86&$-$57.15& 0.33 &24.92&24.82&0.25&24.95\\ 
 &         &1.27&22.25&22.43&24.35&0.17&0.87&$-$56.49& 0.33 &24.92&25.15&0.25&24.88\\ 
 31898&30& disk    &23.15&23.47&24.66&0.33&0.58&13.64& 0.32 &24.53&24.64&0.34&24.95\\ 
 &         &deVauc   &23.08&23.57&24.18&0.57&0.58&14.08& 0.32 &24.53&24.30&0.29&24.73\\ 
 &         &1.71&23.13&23.50&24.49&0.40&0.58&14.16& 0.32 &24.53&24.51&0.34&25.00\\ 
 15731&18& disk    &19.81&19.90&22.67&0.07&0.94&87.59& 3.27 &22.99&21.84&0.11&20.42\\ 
{\it mask}&&deVauc   &19.74&19.89&21.97&0.14&0.88&$-$3.49& 3.29 &22.99&21.88&0.16&21.59\\ 
 &         &0.82&19.82&19.90&22.74&0.07&0.92&85.02& 3.27 &22.99&21.85&0.11&20.43\\ 
 50997&85& disk    &20.66&20.69&24.81&0.02&0.82&$-$13.19& 0.77 &50&25.71&0.29&25.60\\ 
{\it mask}&&deVauc   &20.62&20.63&34.61&$<$0.01&0.11&$-$54.96& 0.73 &50&28.22&0.16&25.77\\ 
 &         &10.00&20.52&20.53&50.00&$<$0.01&0.10&35.78& 0.73 &50&25.41&0.07&21.96\\ 
 50997&51& disk    &20.43&20.46&24.41&0.02&0.67&10.88& 0.63 &24.49&25.04&0.43&25.30\\ 
{\it mask}&&deVauc   &20.42&20.47&23.94&0.04&0.54&1.20& 0.64 &24.49&24.38&0.34&24.83\\ 
 &         &0.70&20.43&20.46&24.52&0.02&0.69&14.08& 0.63 &24.49&25.01&0.38&24.99\\ 
 49298&52& disk    &19.87&20.01&22.21&0.13&0.53&$-$80.46& 2.08 &21.81&22.30&0.70&23.96\\ 
 &         &deVauc   &19.76&20.01&21.52&0.24&0.56&$-$81.40& 2.16 &21.81&22.19&0.70&24.18\\ 
 &         &0.79&19.87&20.00&22.29&0.12&0.52&$-$80.74& 2.08 &21.81&22.44&0.79&24.13\\ 
 49298&85& disk    &19.92&20.03&22.48&0.10&0.41&32.97& 1.50 &23.63&23.50&0.79&25.48\\ 
 &         &deVauc   &19.77&20.05&21.42&0.28&0.36&36.29& 1.52 &23.63&23.34&0.79&25.55\\ 
 &         &0.32&19.92&20.00&22.80&0.07&0.41&29.88& 1.49 &23.63&22.83&0.25&23.79\\ 
 43151&95& disk    &22.30&22.32&26.95&0.01&0.89&81.11& 0.38 &29.82&27.15&0.43&28.58\\ 
 &         &deVauc   &22.25&22.26&35.99&$<$0.01&0.10&46.05& 0.37 &29.82&29.89&0.07&26.45\\ 
 &         &7.66&22.21&22.22&36.23&$<$0.01&0.10&39.37& 0.37 &29.82&50&0&0.\\ 
 43151&45& disk    &22.37&22.42&25.94&0.03&0.88&$-$51.65& 0.39 &28.83&25.96&0.43&26.78\\ 
 &         &deVauc   &22.36&22.41&25.89&0.04&0.37&69.94& 0.39 &28.83&26.15&0.38&27.18\\ 
 &         &0.06&22.51&22.54&26.66&0.02&0.59&71.96& 0.45 &28.83&26.74&0.43&26.96\\ 
\tableline
\end{tabular}

We include for each object: (1) its \combo~ID; (2) the
  \gems\ ~tile of the analyzed image (\goods\ tile when larger than 78); (3) the fitted model, indicated as 
{\it disk} for a exponential profile, {\it deVauc} for a
  Vaucouleurs profile and index {\it N} for a S\`ersic profile, where this
  index shows the best fitted S\`ersic index; (4,5,6) the total, nuclear and host 
  magnitude from the fitting; (7) the host-to-nucleus flux ratio; (8) the
  semiaxis ratio; (9) the position angle; (10) the fit reduced $\chi^2$; (11) 
  the peak subtracted magnitude (not model dependent); (12) the integrated magnitude over the
  restored host image; (13) the half-light radius, in arcsec  and (14) the
  surface brightness at this radius, in mag~arcsec$^{-2}$. If an object image
  has been masked prior to fit we have indicated it with a {\it mask} in colum
  (1).
%

\end{table*}

\begin{table*}
\footnotesize
\caption{Results from the 2D fitting: F850LP-band images}
\label{ltable_z}
\begin{tabular}{ccccccccrccccc}
\tableline\tableline
QSO ID&Tile&Model&Tot&Nuc&Host&H:N&a/b&P.A.&$\chi^2_{\nu}$&Sub&Emp&r$_{\rm 1/2}$&$\mu_{\rm 1/2}$\\ 
\tableline
 34357&94& disk    &18.34&18.66&19.86&0.33&0.87&$-$7.81& 0.59 &19.84&19.96&0.52&21.17\\ 
 &         &deVauc   &18.32&18.81&19.44&0.55&0.88&$-$5.38& 0.57 &19.84&19.66&0.43&20.93\\ 
 &         &3.33&18.34&18.82&19.48&0.54&0.86&$-$4.63& 0.58 &19.84&19.64&0.38&20.64\\ 
 41310&47& disk    &22.56&23.29&23.35&0.94&0.41&39.73& 0.31 &23.30&23.35&0.25&22.99\\ 
 &         &deVauc   &22.53&24.12&22.82&3.31&0.99&$-$4.57& 0.32 &23.30&22.82&0.20&22.45\\ 
 &         &1.57&22.56&23.48&23.18&1.31&0.42&40.08& 0.31 &23.30&23.17&0.25&22.94\\ 
 41310&40& disk    &22.53&23.37&23.22&1.14&0.40&43.08& 0.37 &22.40&23.20&0.25&22.86\\ 
 &         &deVauc   &22.44&24.26&22.67&4.32&0.42&42.29& 0.37 &22.40&22.68&0.20&22.28\\ 
 &         &2.92&22.49&24.18&22.75&3.73&0.44&42.50& 0.37 &22.40&22.75&0.20&22.27\\ 
 52963&50& disk    &22.23&23.41&22.68&1.95&0.56&2.14& 0.36 &22.72&22.66&0.25&22.21\\ 
 &         &deVauc   &22.10&24.20&22.27&5.91&0.47&5.31& 0.35 &22.72&22.28&0.20&21.71\\ 
 &         &3.00&22.14&25.63&22.19&$>$10&0.55&4.73& 0.35 &22.72&22.19&0.20&21.63\\ 
 36361&94& disk    &22.64&25.21&22.75&9.63&0.56&$-$6.03& 0.43 &23.28&22.85&0.25&22.37\\ 
{\it mask}&&deVauc   &22.32&50.00&22.33&$>$10&0.55&$-$8.61& 0.45 &23.28&22.50&0.25&22.42\\ 
 &         &0.95&22.64&25.10&22.76&8.62&0.56&$-$5.93& 0.43 &23.28&22.87&0.25&22.38\\ 
 47615&46& disk    &21.41&22.64&21.84&2.08&0.88&$-$87.85& 0.47 &21.96&21.82&0.30&21.81\\ 
 &         &deVauc   &21.27&23.80&21.39&9.20&0.89&$-$84.30& 0.49 &21.96&21.42&0.25&21.52\\ 
 &         &1.10&21.40&22.68&21.81&2.22&0.88&$-$87.58& 0.47 &21.96&21.79&0.29&21.81\\ 
 50415&85& disk    &21.79&23.12&22.18&2.37&0.87&10.26& 0.46 &22.21&22.24&0.34&22.56\\ 
{\it mask}&&deVauc   &21.58&23.69&21.75&5.97&0.86&9.87& 0.47 &22.21&21.92&0.34&22.69\\ 
 &         &1.09&21.79&23.15&22.16&2.48&0.87&10.21& 0.46 &22.21&22.23&0.34&22.57\\ 
 50415&51& disk    &21.79&23.02&22.22&2.08&0.89&71.82& 0.33 &22.23&22.21&0.34&22.53\\ 
{\it mask}&&deVauc   &21.59&23.66&21.77&5.70&0.88&74.96& 0.34 &22.23&21.85&0.34&22.63\\ 
 &         &1.31&21.76&23.11&22.14&2.44&0.89&72.99& 0.33 &22.23&22.14&0.34&22.56\\ 
 44126&42& disk    &22.35&22.60&24.10&0.25&0.88&$-$20.59& 0.44 &24.68&24.08&0.30&24.09\\ 
 &         &deVauc   &22.36&22.56&24.31&0.19&0.60&$-$25.76& 0.44 &24.68&24.34&0.25&24.29\\ 
 &         &0.16&22.36&22.50&24.67&0.13&0.69&$-$49.28& 0.43 &24.68&24.66&0.29&24.16\\ 
 42601&42& disk    &20.32&22.28&20.52&5.05&0.77&24.14& 0.47 &20.46&20.53&0.56&21.94\\ 
 &         &deVauc   &19.90&23.40&19.95&$>$10&0.78&23.41& 0.44 &20.46&20.32&0.56&22.09\\ 
 &         &2.29&20.13&22.83&20.23&$>$10&0.78&23.99& 0.43 &20.46&20.38&0.56&22.03\\ 
 48284&85& disk    &18.86&19.15&20.47&0.29&0.88&$-$51.18& 0.64 &20.49&20.58&0.52&21.80\\ 
 &         &deVauc   &18.79&19.24&19.98&0.50&0.89&$-$56.14& 0.64 &20.49&20.26&0.47&21.73\\ 
 &         &3.31&18.89&19.31&20.15&0.46&0.95&$-$6.09& 0.71 &20.49&20.30&0.38&21.30\\ 
 39432&89& disk    &21.41&21.87&22.57&0.52&0.94&$-$9.41& 0.40 &22.59&22.69&0.30&22.66\\ 
 &         &deVauc   &21.32&21.93&22.26&0.73&0.90&$-$19.89& 0.41 &22.59&22.40&0.29&22.88\\ 
 &         &1.79&21.41&22.00&22.36&0.71&0.93&$-$43.26& 0.40 &22.59&22.42&0.25&22.28\\ 
 31898&30& disk    &21.86&23.35&22.19&2.91&0.58&17.14& 0.31 &22.01&22.18&0.34&22.44\\ 
 &         &deVauc   &21.61&24.31&21.71&$>$10&0.58&17.61& 0.31 &22.01&21.80&0.34&22.50\\ 
 &         &2.52&21.72&23.93&21.88&6.60&0.58&17.45& 0.30 &22.01&21.91&0.34&22.48\\ 
 15731&18& disk    &19.50&20.02&20.56&0.60&0.99&$-$78.40& 0.77 &20.66&20.74&0.56&22.04\\ 
{\it mask}&&deVauc   &19.30&20.11&20.00&1.10&0.91&$-$24.68& 0.77 &20.66&20.39&0.47&21.83\\ 
 &         &1.83&19.45&20.07&20.36&0.76&0.93&$-$38.40& 0.76 &20.66&20.45&0.47&21.78\\ 
 50997&85& disk    &20.36&20.49&22.75&0.12&0.63&$-$23.78& 0.49 &22.94&22.16&0.20&21.81\\ 
{\it mask}&&deVauc   &20.34&20.65&21.88&0.32&0.80&$-$15.77& 0.52 &22.94&22.01&0.25&22.09\\ 
 &         &0.83&20.36&20.58&22.22&0.22&0.79&$-$13.44& 0.52 &22.94&22.27&0.29&22.22\\ 
 50997&51& disk    &20.29&20.40&22.84&0.10&0.63&40.80& 0.39 &22.77&22.85&0.29&22.85\\ 
{\it mask}&&deVauc   &20.27&20.46&22.29&0.18&0.61&44.89& 0.39 &22.77&21.83&0.16&20.88\\ 
 &         &1.39&20.28&20.41&22.73&0.11&0.62&41.68& 0.39 &22.77&22.75&0.29&22.85\\ 
 49298&52& disk    &19.46&19.79&20.92&0.35&0.55&$-$82.13& 0.79 &21.13&20.78&0.65&22.65\\ 
 &         &deVauc   &19.13&19.82&19.96&0.87&0.47&$-$81.49& 0.82 &21.13&21.07&0.88&23.24\\ 
 &         &0.95&19.46&19.79&20.94&0.34&0.55&$-$82.22& 0.79 &21.13&20.78&0.65&22.64\\ 
 49298&85& disk    &19.50&19.83&20.99&0.34&0.50&28.65& 0.72 &21.10&20.86&0.56&22.61\\ 
 &         &deVauc   &19.24&19.86&20.15&0.76&0.45&30.25& 0.75 &21.10&21.09&0.75&23.23\\ 
 &         &0.70&19.51&19.81&21.09&0.30&0.50&27.75& 0.72 &21.10&20.93&0.56&22.58\\ 
 43151&95& disk    &21.80&22.12&23.31&0.33&1.00&68.38& 0.51 &23.31&23.39&0.20&22.62\\ 
 &         &deVauc   &21.75&22.23&22.87&0.55&0.83&68.29& 0.51 &23.31&22.80&0.16&21.85\\ 
 &         &0.87&21.80&22.10&23.36&0.31&0.83&71.44& 0.51 &23.31&23.11&0.20&22.51\\ 
 43151&45& disk    &21.93&22.22&23.52&0.30&0.98&82.06& 0.39 &23.72&23.55&0.29&23.52\\ 
 &         &deVauc   &21.91&22.37&23.07&0.52&0.89&$-$79.39& 0.39 &23.72&23.08&0.25&23.20\\ 
 &         &1.20&21.92&22.23&23.44&0.32&0.97&79.46& 0.39 &23.72&23.44&0.29&23.50\\ 
\tableline
\end{tabular}

%
Columns as in table \ref{ltable_V}
%

\end{table*}

 \section{Results}\label{sec:results}
 
 Based on our fitting analysis and the simulations, we assigned a
 morphological class to each of the AGN hosts (Tables \ref{ltable_V} and
 \ref{ltable_z}).  We selected the morphological type for each host galaxy by
 comparing its location in the S\`ersic index vs.\ host magnitude plane with
 the distribution obtained from the simulations (Fig.\ \ref{sersic}).  This
 way, we derived initial morphological classifications for each filter band
 separately. If an object was observed more than once, in different mosaic
 tiles (see Tables \ref{ltable_V} and \ref{ltable_z}), we used the average
 value of the two S\`ersic indices. If one of the two images originated in the
 \goods\ area, we gave preference to the values derived from the analysis of
 the \gems\ tile (since the \goods\ tiles incorporated into \gems\ are
 slightly shallower that the proper \gems\ tiles, cf.\ Sect.~\ref{sec:data}).
 From these criteria we could obtain unambiguous classifications for 11 of the
 15 objects, of which 9 have the same type in both bands, and 2 were detected
 only in F850LP.  All these 11 objects are located in a non-overlapping region
 between the spheroid and disk distributions of Fig.\ \ref{sersic}.  For 3 of
 the remaining 4 objects, the classification was clear-cut in one band, and
 ambiguous in the other band (\combo\ 36361 and 49298 in F606W, \combo\ 50415
 in F850LP). In these cases we adopted the morphological classification from
 that band where the type was found to be unambiguous.  Only for one remaining
 object we could not converge on a unique classification based on the fits
 alone. This object (\combo\ 15731) has a S\`ersic index of 0.82 in F606W,
 indicative of a disk, and of 1.83 in F850LP, indicative of a spheroid.  After
 visual inspection of images and surface brightness profiles we concluded that
 this object is most likely an early-type, bulge-dominated galaxy (expecially
 in the redder F850LP band) without significant evidence of a disk, but with
 two faint blue arms sticking out in the F606W (rest-frame $B$ band) residual
 images (see Fig.\ \ref{maps}).  These ``arms'' could also be tidal features
 resulting from interaction with the close companion of that galaxy, owing
 their blue colors to enhanced star formation. We also note that contrast
 between nucleus and host is considerably more favourable in the F850LP band
 for this $z=0.83$ AGN, giving additional weight to its classification as an
 early-type galaxy.

 Based on the assigned type, we estimated the nuclear magnitude using the
 value derived from the 2D fitting. The host magnitude was measured from the
 actual data through a large circular aperture after removing the nucleus.
 The structural parameters $r_{1/2}$ and $\mu_{1/2}$ were obtained from the
 surface brightness analysis. Finally, we flagged each object with a value
 between 0 and 3 to describe its degree of interaction (based upon visual
 inspection), where 0 means no nearby companion within a projected radius of
 $\sim 2\arcsec$ (i.e., $\sim$13~kpc at the mean redshift of our sample); 1
 means at least one nearby companion, but no tidal tails or bridges between
 the host and the companion; 2 indicates a nearby companion with an apparent
 bridge between the host and the companion, and 3 signifies objects clearly
 undergoing a merger event.  Table \ref{final_table} lists the final estimated
 parameters for our objects, separately for the F606W and F850LP-band images,
 including the morphological classification and the interaction level flag.
 The listed magnitudes have not been corrected for Galactic extinction, which
 is however small. For the values quoted hereafter, we adopt a Galactic
 extinction of $E(B-V)=0.008$~mag, which corresponds to corrections of
 $\sim$0.024 and $\sim$0.014~mag for the F606W and F850LP bands, respectively
 \citep{schl98}.  In the Appendix we present color images and surface
 brightness profile of the original data, the model, the restored host galaxy
 and the residuals, together with notes on individual objects.

\begin{figure}
\includegraphics[angle=-90,width=\hsize]{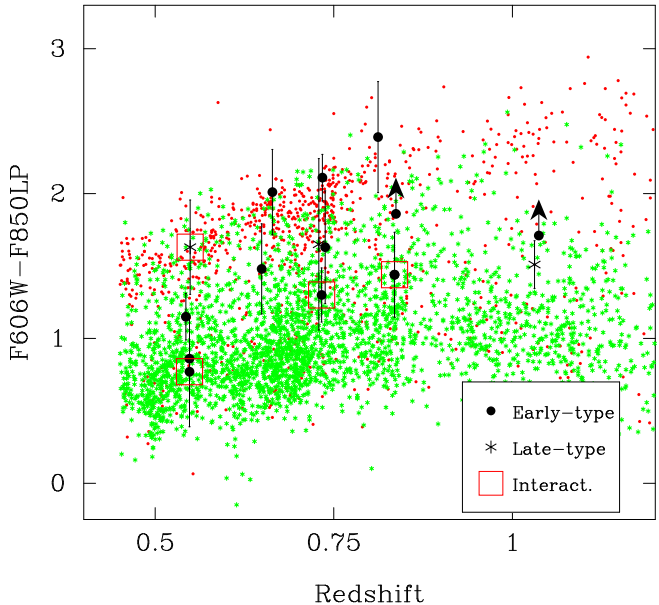}
\caption{\label{VZ_z} 
  Observed $\V - \Z$ colors for the host galaxies as a function of redshift
  (large symbols).  The solid circles show morphologically early-type
  galaxies, the stars morphologically late-type galaxies (see text for
  details). The open squares mark the objects with evidence of
  interactions/mergers (class 2 and 3 in Table \ref{final_table}). The small
  symbols indicate early-type ($n\ge 2.5$, red solid circles) and late-type
  ($n<2.5$, green stars) inactive galaxies observed in the \gems\ field.
}
\end{figure}

\begin{figure}
\includegraphics[angle=-90,width=\hsize]{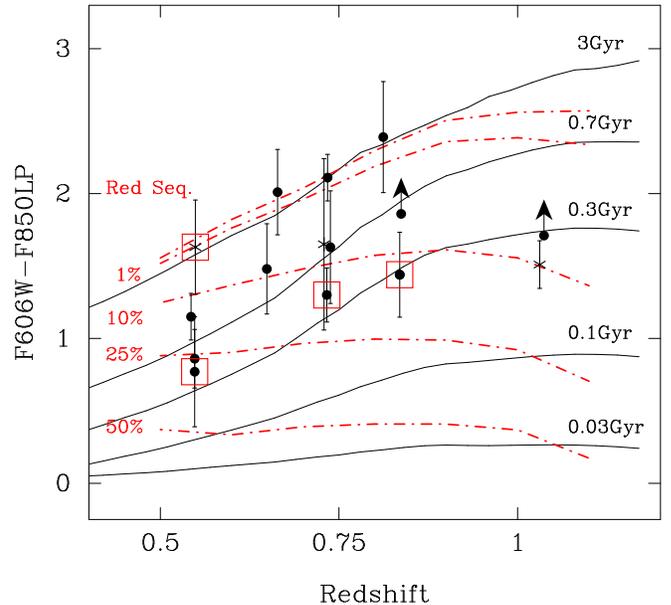}
\caption{\label{VZ_z_tr} 
  Observed $\V - \Z$ colors for the host galaxies as a function of redshift
  (large symbols), compared to simple populations models.  The symbols are the
  same as those in Fig.~\ref{VZ_z}.  The solid lines show the expected colors
  of a single stellar population with different ages.  The dashed lines show
  the expected colors of a mixed population, with a dominant old stellar
  population and a varying contribution of light from young stellar
  populations, derived completely empirically (see Section 5.1 for details).
}
\end{figure}

 We find that 3 of the 15 objects show evidence of a major merger: \combo\ 
 52963, \combo\ 36361 and \combo\ 42601. The images of these objects show
 generally blue clumpy structures, with linear shapes, as expected from star
 forming regions induced by recent merger events. In some cases it is
 difficult to distinguish these structures from those expected from spiral
 arms (e.g., \combo\ 42601), although the lack of disk-structure in the
 profile suggests a merger process. Another object, \combo\ 15731 shows a
 companion with what seems to be a bridge connecting it with the host.
 Finally, two other objects show what could be close companions, without any
 evidence of connection between them and the hosts. Therefore, a fraction
 between 20\% and 40\% of the objects shows some degree of interaction with
 close companions, from major merger to low-level interaction.
 
 Twelve of the 15 hosts -- $\sim$80\% of the sample -- are morphologically 
 early-type galaxies, as defined by the S\`ersic index. This fraction is not
 affected if we exclude the 3 possibly merging galaxies.  On the other hand,
 only one object can be clearly classified as a disk-dominated galaxy (\combo\ 
 49298). Two other galaxies could also have significant disks, of which 
 \combo\ 36361 is heavily distorted, and the classification is uncertain.

\begin{table*}
\footnotesize
\caption{Summary of the Results of our analysis}
\label{final_table}
\begin{tabular}{cccccccclc}
\tableline\tableline
QSO ID&$z$&Class&F606W$_{\rm QSO}$&F606W$_{\rm Nuc}$&F606W$_{\mathrm{host}}$&r$_{\rm  1/2}$&$\mu_{\rm 1/2}$&Morph.&Inter.\\
\tableline
34357&0.543&deVauc  &19.19$\pm$0.02&19.53$\pm$0.03&20.82$\pm$0.05&2.73$\pm$0.72&21.4$\pm$0.2& Early Type & 0\\
41310&0.548&deVauc  &23.24$\pm$0.04&24.58$\pm$0.06&23.55$\pm$0.26&0.74$\pm$0.23&20.3$\pm$0.2& Early Type & 1\\
52963&0.548&deVauc  &22.78$\pm$0.03&25.05$\pm$0.07&22.97$\pm$0.22&0.68$\pm$0.21&19.4$\pm$0.2& Early Type & 3\\
36361&0.549& disk   &23.39$\pm$0.02&25.28$\pm$0.07&24.24$\pm$0.14&1.32$\pm$0.42&22.1$\pm$0.2& Late Type & 3\\
47615&0.649&deVauc  &22.07$\pm$0.03&22.77$\pm$0.03&22.91$\pm$0.17&1.12$\pm$0.34&20.4$\pm$0.2& Early Type & 0\\
50415&0.664&{\it deVauc}&22.96$\pm$0.03&23.45$\pm$0.03&24.00$\pm$0.14&2.13$\pm$0.61&22.2$\pm$0.2& Early Type& 1\\
44126&0.729& disk   &23.07$\pm$0.04&23.16$\pm$0.06&24.46$\pm$0.52&0.63$\pm$0.17&20.7$\pm$0.2& Late Type  & 0\\
42601&0.733&deVauc  &20.83$\pm$0.02&23.25$\pm$0.05&21.59$\pm$0.06&3.75$\pm$1.15&21.2$\pm$0.3& Early Type & 3\\
48284&0.734&deVauc  &19.36$\pm$0.02&19.53$\pm$0.03&22.38$\pm$0.05&1.85$\pm$0.58&20.7$\pm$0.2& Early Type & 0\\
39432&0.738&deVauc  &22.19$\pm$0.02&22.44$\pm$0.03&24.73$\pm$0.26&1.44$\pm$0.46&21.7$\pm$0.3& Early Type & 0\\
31898&0.812&deVauc  &23.08$\pm$0.04&23.57$\pm$0.06&24.20$\pm$0.26&1.89$\pm$0.59&22.3$\pm$0.3& Early Type & 0\\
15731&0.835& disk   &19.81$\pm$0.02&19.90$\pm$0.03&21.84$\pm$0.17&0.65$\pm$0.18&17.8$\pm$0.1&{\it Early Type}& 2\\
50997&0.837&Not Det.&20.62$\pm$0.02&20.63$\pm$0.03& ---          & ---         & ---        & Early Type & 1\\
49298&1.031& disk   &19.87$\pm$0.02&20.01$\pm$0.03&22.30$\pm$0.05&4.27$\pm$1.28&21.9$\pm$0.2& Late Type  & 0\\
43151&1.037&Not Det.&22.30$\pm$0.02&22.32$\pm$0.03& ---          & ---         & ---        & Early Type & 0\\
\\
\tableline\tableline
QSO ID&$z$&Class&F850LP$_{\rm QSO}$&F850LP$_{\rm Nuc}$&F850LP$_{\mathrm{host}}$&r$_{\rm 1/2}$&$\mu_{\rm 1/2}$\\ 
\tableline
34357&0.543&deVauc   &18.32$\pm$0.02&18.81$\pm$0.05&19.66$\pm$0.06&2.15$\pm$0.63&18.9$\pm$0.2\\ 
41310&0.548&deVauc   &22.53$\pm$0.05&24.02$\pm$0.16&22.67$\pm$0.37&1.16$\pm$0.35&20.6$\pm$0.3\\ 
52963&0.548&deVauc   &22.10$\pm$0.04&24.10$\pm$0.16&22.19$\pm$0.37&1.05$\pm$0.32&19.8$\pm$0.3\\ 
36361&0.549&{\it disk }   &22.64$\pm$0.05&25.13$\pm$0.19&22.60$\pm$0.38&1.32$\pm$0.42&20.5$\pm$0.3\\ 
47615&0.649&deVauc   &21.27$\pm$0.04&23.71$\pm$0.15&21.42$\pm$0.22&1.51$\pm$0.45&19.5$\pm$0.3\\ 
50415&0.664&deVauc   &21.58$\pm$0.04&23.60$\pm$0.15&21.79$\pm$0.26&1.91$\pm$0.56&20.4$\pm$0.3\\ 
44126&0.729& disk    &22.35$\pm$0.04&22.60$\pm$0.07&22.80$\pm$0.60&1.67$\pm$0.52&21.7$\pm$0.5\\ 
42601&0.733&deVauc   &19.90$\pm$0.02&23.40$\pm$0.13&20.28$\pm$0.14&3.23$\pm$0.98&19.7$\pm$0.3\\ 
48284&0.734&deVauc   &18.79$\pm$0.02&19.24$\pm$0.05&20.26$\pm$0.06&2.48$\pm$0.75&19.1$\pm$0.3\\ 
39432&0.738&deVauc   &21.32$\pm$0.04&21.82$\pm$0.14&22.30$\pm$0.29&1.56$\pm$0.46&20.2$\pm$0.3\\ 
31898&0.812&deVauc   &21.61$\pm$0.04&24.21$\pm$0.16&21.80$\pm$0.26&2.00$\pm$0.59&19.9$\pm$0.3\\ 
15731&0.835&deVauc   &19.30$\pm$0.02&20.11$\pm$0.05&20.39$\pm$0.06&2.79$\pm$0.82&19.2$\pm$0.3\\ 
50997&0.837&deVauc   &20.34$\pm$0.02&20.65$\pm$0.05&21.92$\pm$0.26&1.36$\pm$0.40&19.3$\pm$0.3\\ 
49298&1.031&{\it disk} &19.46$\pm$0.02&19.79$\pm$0.05&20.78$\pm$0.09&3.97$\pm$1.22&19.6$\pm$0.2\\ 
43151&1.037&deVauc   &21.75$\pm$0.04&22.12$\pm$0.14&22.65$\pm$0.36&1.65$\pm$0.46&20.3$\pm$0.3\\ 
\tableline
\end{tabular}

%
We include for each object and band the
\combo~id. (1), the redshift (2), the host classification (italics when dubious) (3), the total
magnitude (4), the nucleus and host magnitudes (5,6), the host structural
 parameters, the half-light radius in kpc (7) and the surface brightness at
 this radius, corrected for cosmological dimming (8), 
the final morphological classification (italics when dubious) (9) and a flag indicating the interaction degree (10). 
The interaction degree has been visually classified as: {\it 0} Isolated galaxies,
{\it 1} Galaxies with a close companion, {\it 2} Galaxies with tidal tails and bridges
with a close companion and {\it 3} clear mergers.
The listed magnitudes have not been corrected for galactic extinction.
%

\end{table*}

\section{Discussion}\label{sec:discussion}

\subsection{The colors of the host galaxies}

The average observed color of all the hosts is $\V - \Z =1.55\pm 0.45$~mag,
without significant differences between morphological classes ($1.54\pm 0.48$
for the early-type hosts and 1$.65\pm 0.09$ for the late-type hosts, excluding
mergers and strongly interacting systems).
We have compared the host and nuclear colors in order to check if there were a
possible contamination from the nucleus, due to improper subtraction, that
could affect our results. No significant trend is seen between the nuclear and
host colors.

Figure \ref{VZ_z} shows the F606W$-$F850LP distribution of the host galaxies
as a function of redshift. For comparision, we included the same distribution
for the $\sim$4000 inactive galaxies in the \gems\ field at the redshift range
of our objects.  A detailed study of these objects will be presented in
forthcoming papers by the \gems\ collaboration (Barden et al., in prep.; Wolf
et al., in prep.; McIntosh et al., in prep.).  The red sequence of early-type
galaxies, as discussed by \cite{bell04}, is clearly identified in the figure
as a red envelope, and shows the color of the oldest stellar populations at a
give epoch. It is clear from Fig. \ref{VZ_z} that the host galaxies, despite
their morphological types, range in their colors from the red sequence to blue
colors indicative of ongoing or recent star formation.  In particular, the
morphologically early-type hosts tend to be bluer, on average, than the red
sequence galaxies, hinting at the presence of significant younger stellar
population.  Roughly $50-70$\% of the 10 early-type hosts detected in both
bands are bluer than the red sequence galaxies.

There are several combinations of stellar populations that could explain the
observed distribution. We discuss two extreme cases: (a) The galaxy
has a single stellar population, formed in a single burst of star formation;
and (b) the galaxy is dominated by an old stellar population, similar to the
stellar population of field early-type galaxies, but it has undergone recent
star formation.  Figure \ref{VZ_z_tr} shows schematically the same color 
distribution as Fig. \ref{VZ_z}, but now we overplotted the expected colors 
for the two scenarios. Solid lines show the colors of single stellar populations. 
These colors were calculated using the \cite{bruz03} models, assuming a solar
metallicity and a \cite{chab03} IMF (we also tried a \cite{salp55}
IMF, without significant differences). These isochrones are a rough indicator
of the luminosity-weighted age of the host galaxies. A wide range of ages,
from $\sim$0.3~Gyr to $\sim$3~Gyr, is required to explain the observed
distribution. In average they seem to have moderately young
luminosity-weighted stellar populations, with an age of $\sim$1~Gyr.  These
ages are indicators of the wide range of stellar populations needed to
describe the colors of the host galaxies in our sample.

Yet, a mix of different populations can also generate similar colors.  The
dashed lines in Figure \ref{VZ_z_tr} show the colors of \gems\ red sequence
galaxies (taken to empirically represent an old stellar population), where a
varying fraction of young stellar population has been added to the overall
mass.  The young stellar population has been characterized by the population
of the galaxies in the blue end of the $U-V$ color-diagram distribution for
the GEMS field ($U-V<0$ mag). This model-independent approach has been
cross-checked substituting this young population with a 0.03~Gyr model without
appreciable changes. It is seen that the colors of the host galaxies can be
reproduced by adding a varying mass fraction between close to 0~\% and 
$\sim$30~\% of young stars to an overall population of old stars.

It is worth noting that uncertainties from dust reddening and metallicity only
mildly affect our conclusion that substantial young stellar populations are
present in many of the AGN host galaxies. In particular, dust tends to dim and
redden young stars preferentially \cite[e.g.][]{calz94,zari99}, which, if
corrected for, would imply even larger young stellar population fractions in
our hosts. On the other hand, changes in metallicity by a factor 3, in the
sense that hosts would have lower metallicity, could reproduce the color
offsets seen between the early-type inactive and active galaxies. However,
recalling the typically high luminosities of the AGN host galaxies, coupled
with the metallicity$-$luminosity correlation and inspecting the spatial
clumping of the blue stars in many of the recovered host galaxy images (Fig.
\ref{maps}), it seems rather unlikely that a metallicity difference between
active and non-active morphologically early-type galaxies could be the
dominant factor in driving the observed color differences. Moreover, it is
well known that QSO broad-line regions have high metallicities
\citep{hama92,hama93,cons02}. There is thus no evidence for low metallicities 
in the host galaxies as an explanation for the blue colors.

\begin{figure}
\includegraphics[angle=-90,width=\hsize]{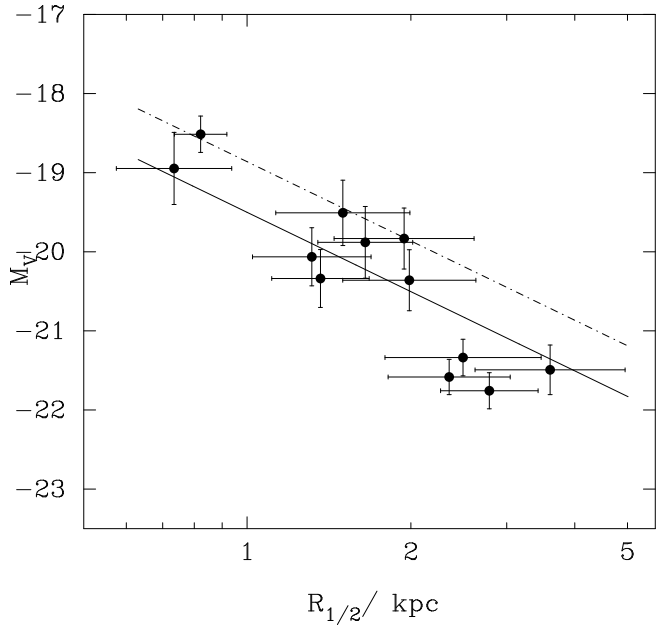}
\caption{\label{Kor} 
  Plot of the host galaxy absolute magnitude against half-light radius for the
  12 early-type galaxies of our sample (solid circles). The solid line shows
  the location of the luminosity-size relation for early-type red sequence
  galaxies at the mean redshift of our objects \citep{scha97}. The dashed-dotted
  line shows the relation at $z=0$ \citep{korm77}.
  }
\end{figure}

\begin{figure*}
\includegraphics[angle=-90,width=\hsize]{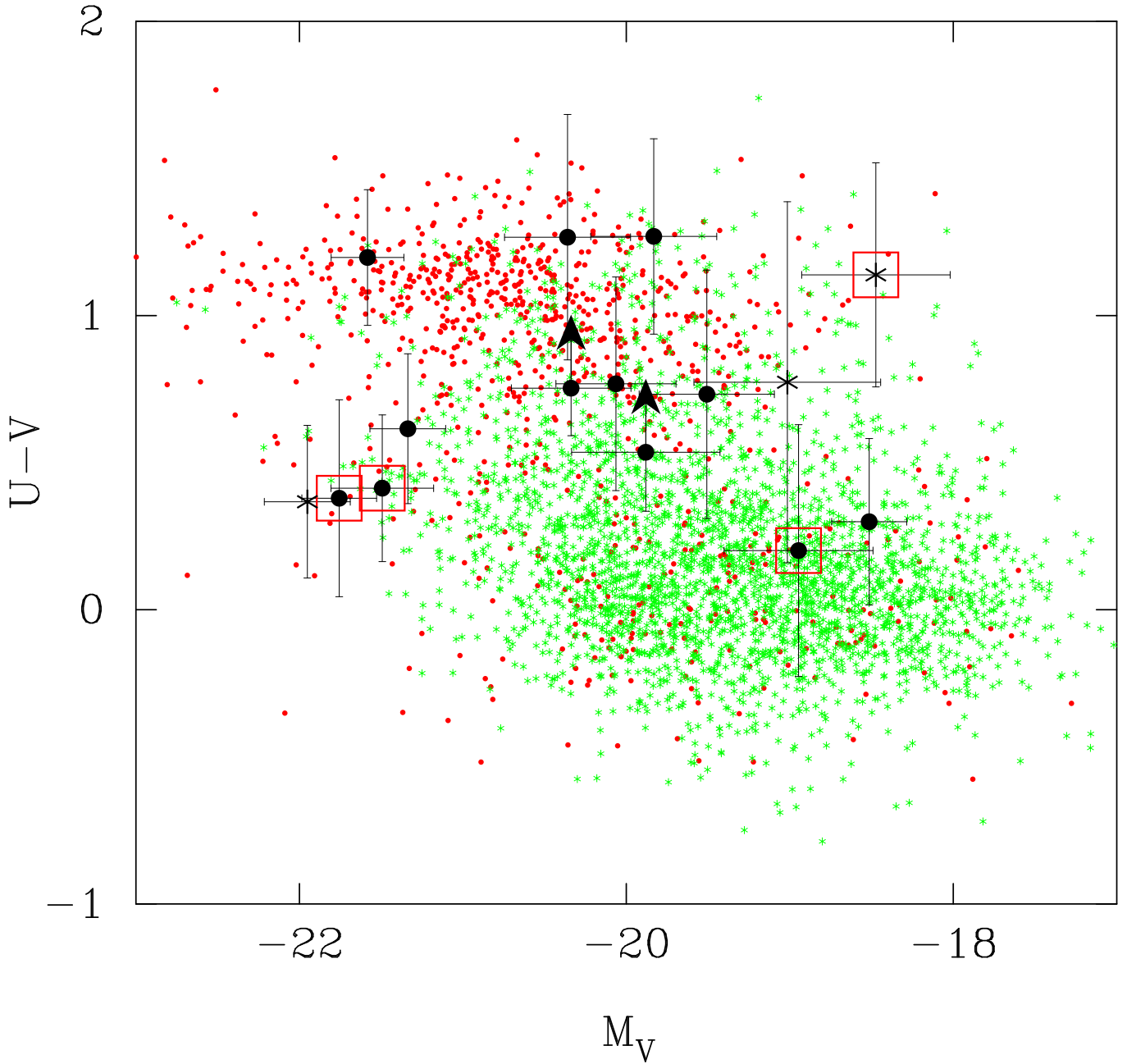}
\caption{\label{VZ_Mz} 
  Rest-frame color-magnitude distribution of the host galaxies of active
  nuclei compared with the inactive galaxies from \gems\ from the same
  redshift range as our AGNs (symbols as in Fig.~\ref{VZ_z}) .  The red
  sequence for early types is clearly identified in the field galaxy
  distribution (at $U-V\ga 0.8$).  The hosts, mainly early-type galaxies, are
  bluer than the red-sequence galaxies in the field.  There is a clear cutoff
  in the upper-right corner due to combined effect of the different detection
  limits in the F606W and the F850LP bands.  This limit does not affect our
  results.
}
\end{figure*}

\subsection{Absolute magnitudes, scale-lengths and stellar masses of the host}

We calculated $k$-corrections for the host galaxies to obtain rest-frame $U-V$
colors and absolute magnitudes that allowed us to compare them with the
inactive galaxies in the \gems\ field. In the Appendix (Section \ref{a2}), we
describe in detail the method adopted to derive those transformations. We used
the observed $\V-\Z$ colors to derive the rest-frame $U-V$ colors, using these
transformations, including the standard deviation of each transformation in
the color errors. Recall that for the redshift range of the sample, these
bandpass shifts from $\V-\Z$ to $U-V$ and therefore the color transformations
are relative small. Once determined the rest-frame $U-V$ colors, we derived
the absolute magnitudes of our objects, using the transformations described in
the Appendix, including the standard deviation of each transformation in the
magnitude errors.  The mean $V$-band absolute magnitudes of our AGN host
galaxies is $M_{V,\mathrm{host}}=-20.2\pm 1.2$ mag, covering a range between
$\sim -18.5$ and $\sim -22$ mag.

We determined the half-light radii of these galaxies from the scale lengths
obtained in the fits, computing a luminosity weighted average of the
values obtained for both bands (Table \ref{final_table}).  The average 
half-light radius is $r_{1/2}\sim 1.9 \pm 0.8$~kpc.  Figure \ref{Kor}
shows the distribution of the absolute magnitudes as function of these radii.
For comparison, we overplot the luminosity-size relation of early-type
galaxies at the average redshift of our sample \citep{korm77,fabe87}. 
We adopted the luminosity-size relation obtained by \citet{scha97} 
for red sequence early-type galaxies in clusters at $z\sim 0.7$, transformed to
our cosmology.  The loci of the early-type active and inactive galaxies in
the luminosity-size plane are consistent with each other.

We estimated the stellar masses using the average relation found by
\cite{bell01} between the $M/L$ and the $B-V$ color. For that we have
estimated the $B-V$ colors of our objects using the relation between the
$U-V$ and the $B-V$ colors derived from the field galaxies in the
\combo.  The correlation between both colors is tight, with a dispersion of
$\sim$0.07 mag. Using these colors and the $V$-band absolute magnitudes of our
objects we have an estimate of the stellar masses of our host galaxies. 
The average mass derived for the early-type hosts is $\sim 1.5\times
10^{10} M_\odot$, with a range of masses between $\sim 0.1$ and $\sim
5.9\times 10^{10} M_\odot$. We find AGN harboured by early-type galaxies
with a wide range of masses, not restricted to the most massive ellipticals,
as previously reported by the studies of more luminous QSO hosts
\cite[e.g.][]{dunl03}.

\subsection{The $U-V$ color-magnitude diagram}

Figure \ref{VZ_Mz} shows the $U-V$ vs. $M_{V}$ color-magnitude diagram
for the hosts, together with these values for $\sim$4000 inactive galaxies
from the \gems\ sample at the same redshift of our objects. For
consistency we have used the same grid of transformations (described above)
for deriving the rest-frame colors and absolute magnitudes of the active and
inactive galaxies, using the observed F606W and F850LP magnitudes. The
morphological early- and late-type galaxies, as classified within the \gems\ 
project \citep[Barden et al., in prep.;][]{bell04}, have been
indicated with red solid circles and green stars, respectively.
The red sequence of early-type galaxies, described by \cite{bell04} for the
\gems\ inactive galaxies at $z\sim$0.7, is clearly visible in Fig.
\ref{VZ_Mz}. From this figure, we can see that between $5-7$ of the
morphological early-type hosts are bluer than the red sequence early-type
galaxies ($\sim 50-70$\% of the sample). The range depends on where we 
draw the line
for including objects in the red sequence ($U-V\sim 0.6-0.8$).
On the other hand, only $\sim$30~\% of the host galaxies detected in both
bands are clearly as red as red sequence galaxies. This fraction would
increase to $\sim$40\% if we included the two hosts not detected in the
F606W-band.

We performed 1D and 2D Kolmogorov-Smirnoff tests on the color and
color-magnitude distributions of the early-type host galaxies and the
early-type inactive red sequence galaxies. The null hypothesis that both samples
were drawn from the same parent population has probabilities of only
0.1 and 0.4~\%, respectively. The standard KS tests do not take
the errors into account. To determine the extend to which these errors can affect
our results we simulated 1000 color and color-magnitude distributions. We add
random values to the original colors and magnitudes following Gaussian
distributions with the width of the errors.  Repeating the KS tests on these
simulated distributions
we find that the acceptance probabilities of the null hypothesis 
are only marginally larger, with 0.8 and 0.5~\%, respectively. We conclude
that the distribution of early-type AGN host galaxies is significantly
different than that of the red sequence.

However, as previously reported by \cite{bell04} there is a tail of 
blue, but morphologically early-type galaxies which is clearly seen Fig.~\ref{VZ_Mz}.
Using a similar KS test, we find the color and
color-magnitude distributions of the early-type host galaxies are actually
consistent with being drawn from the \emph{complete} sample of inactive 
early-type galaxies, i.e.\ including the blue tail.  
The probabilities found by a KS test on both
samples are 16.7~\% (21.7~\%) and 39.1~\% (34.1~\%), for the color and
color-magnitude distributions without and with the inclusion of errors,
respectively. However, applying a luminosity cut at $M_{V}<-21$, which
isolates the four most luminous early-type host galaxies, the color
distributions are again markedly different, although due to the reduce number
of objects a KS test would not produce a reliable probability.
These objects harbor the four most luminous AGNs of our
sample.

In summary, we have found that the host galaxies of AGNs are mainly elliptical
galaxies ($\sim$~80\%), with $20-40$~\% showing evidence of interactions. Only
$\sim$20~\% are disk dominated. Roughly $50-70$~\% of the host galaxies that
are structurally early-type are bluer than the red sequence early-type
galaxies, indicating the presence of younger stellar populations than that of
the inactive early-type galaxies.  Taking into account the recently reported
blue tail for early-type galaxies \citep{bell04}, their colors become more
similar, but they still remain significantly bluer for the most luminous
objects.  However, they are structurally similar to red sequence early-type
galaxies, which may indicate that they are dominated by an underlying old
stellar population with only a small fraction of the mass involved in a recent
star formation episode.

\subsection{Comparison with recent results}

The majority of previous studies of host galaxies was focused on host
morphologies \citep[e.g.][]{dunl03,sanc03b}. They found that the QSO hosts
were mainly early-type galaxies, with an increase of the fraction of
early-type galaxies with nuclear luminosity \citep{dunl03}.  The fraction of
late-type host galaxies ranges between $\sim$15 and $\sim$35~\% for
radio-quiet QSOs, and almost zero for radio-loud QSOs
\citep{dunl03,floy03,sanc03b,hami02}. These fractions are similar to those
found in our sample ($\sim$20\%), and there is no significant change with
redshift.  In contrast, very few studies have focused on colors. \cite{kauf03}
have recently presented their results based in the comparison of active and
inactive galaxies from the SDSS survey at $0.02<z<0.3$.  They found that the
host galaxies of both type 1 and 2 AGNs have younger mean stellar ages than
inactive galaxies, possibly due to starbust or post-starburst populations.
Furthermore, $\sim$40~\% of their objects are blue \emph{pure} spheroids, and
they found evidence of recent interactions in $\sim$30~\% of their objects.
Their subsample of type 1 AGNs covers a similar range of luminosities compared
to our sample. \cite{jahn02a} already noticed that the mean colors of the
hosts of their complete sample of low-$z$ type 1 AGNs were bluer than
expected, especially for pure elliptical galaxies. Similar results have been
found by \cite{koti04}, for a sample of low-$z$ BL Lac host galaxies.  Their
results are remarkably similar to the ones presented here, especially
regarding the existence of a population of blue elliptical hosts.  Our results
extend this result to early epochs, covering the last half of the cosmic
history.

On the other hand, \cite{dunl03} did not find expecially blue colors for their
sample of low-$z$ high luminosity QSO hosts, based on the analysis of $R$ and
$K$-band images \citep{dunl93,mclu99}.  However, the fraction of young stars
required to reproduce our observed colors would not produce significant
deviations in $R-K$ color at low-$z$ because the red light from blue stars
is small compared to the light from the dominant underlying stellar
populations.  Furthermore, \cite{nola01} show evidence for a small fraction of
young stellar population in their spectroscopic analysis of the host galaxies
studied by \cite{dunl03}. 

In a companion article \citep{jahn04a}, we have analyzed the host galaxies of
AGNs in the \gems\ project at high-$z$ ($1.8<z<2.5$).  We found evidence for
significant UV flux from young stars in all the detected host galaxies, in a
clear agreement with current results.

It seems that the basic ingredients in generating a powerful AGN are a massive
black hole \emph{and} an abundant fuel supply, as already expressed by \cite{kauf03}.
Only massive early-type galaxies contain massive black$-$holes, and only
galaxies with significant amounts of recent star formation have the required
fuel supply. It is still an open question why these early-type galaxies retain
enough gas to undergo star formation, and what is the relationship between the
AGN and the star formation activity.

\begin{figure}
\includegraphics[angle=-90,width=\hsize]{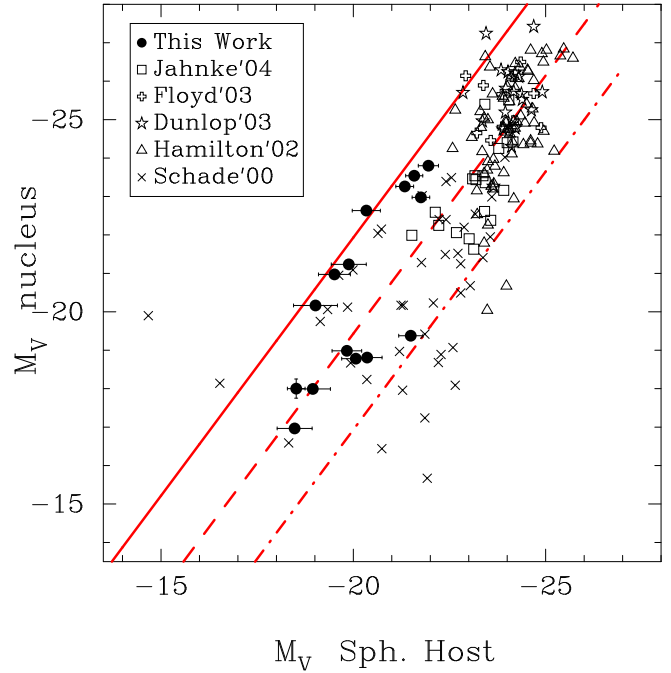}
\caption{\label{HN2} 
  Distribution of the F606W-band absolute magnitudes of nucleus along the
  F606W-band absolute magnitude of the spheroidal host galaxies of our sample
  (solid circles), together with different samples taken from the literature.
  The solid, dashed and dot-dashed lines show the expected location for an AGN
  emitting at 100~\%, 10~\% and 1~\% of the Eddington limit, respectively.
}
\end{figure}

\subsection{AGN activity and star formation: why are our early-type hosts blue?}

The correlation between nuclear activity and enhanced star formation could
arise from an enrichment of gas due to recent merger events. The merger would
produce the infall of gas needed to (re)ignite the AGN and trigger star
formation at the same time \cite[e.g.,][]{lin88,miho96,corb00}. We find
evidence of recent mergers in $\sim$20~\% of our sample, a fraction similar to
the values reported in the literature \cite[e.g.,][]{kauf03,sanc03b}.
However, it is not clear that this fraction could be considered as a
significant excess of merging galaxies. The fraction of merging galaxies in
the local universe is rather small, $\sim$7\%, based on classical studies like
\cite{shap32}. But this fraction is known to increase with the redshift.  For
example, \cite{berg96} and \cite{lee00} found that at least $\sim$39~\% of the
galaxies in the HDF-N and HDF-S show disrupted morphologies, reminiscent of
mergers.
Therefore, if a merger is the cause of the enrichment of gas in these
galaxies, this event must have happened long before the morphological
relaxation of the galaxy, in the majority of the cases.

Galaxy interactions have been claimed for decades to be important triggers of
star formation \citep[e.g.]{lars78,kenn87}. It was also belived that mergers
produce an enrichment of gas, and even minor interactions can trigger star
formation due to the infall of gas, compression and heating
\citep[e.g.][]{bart00}. In particular, tidal interaction can trigger star
formation that significantly affects the colors of the nuclear region
\citep[e.g.][]{bart03}.  \cite[But see ][ for a counter example]{berg03}.
This may well be the kind of effect we detect in our observations, where we
can reproduce the observed colors with a small fraction of young stars.
\cite{miho96} estimated the timescales of star formation induced by 
interaction  to be $\sim 10-100$ Myr, which is of the same order as the
morphological relaxation time of $\sim 50-100$ Myr (depending on the scale length
of the galaxies involved). Since even a somewhat aged population of
recently formed stars can still produce substantially blue colors in the host
galaxy (see the model tracks in Fig. \ref{VZ_z_tr}), we may well still see the
enhanced blue color while the morphological signatures of the merger or
interaction have disappeared.

Another possibility could be that the gas was already present in the
morphological early-type parent galaxy. As stated above, there is a tail of
blue elliptical galaxies in the color distribution of early-type galaxies in
the \gems\ sample -- that is, morphological early-type galaxies with enough 
gas to undergo star formation. In that case a process such as a minor merger, 
which would hardly detectable by our current morphological analysis, 
could produce the infall of this gas into the inner regions. 

A completely different connection between AGN and star formation activity
could come from AGN-driven feedback processes.  The combined effect of the
ionizing radiation and possible mechanical outflows trigger star formation
throughout the host galaxy. The implications of AGN feedback for galaxy 
evolution in general are still poorly understood \citep[e.g.][]{colb96,silk98}.

\subsection{The host-nucleus luminosity relation}

\cite{mcle94a} found that a minimum host luminosity appears to be required to
produce an AGN of given luminosity.  This result has been confirmed by
different authors \citep[e.g.,][]{dunl03, floy03, sanc03b, jahn02a}.
Therefore, the upper envelope of the nuclear luminosities shows a correlation
with host luminosities. The widely accepted interpretation is that for a given
host luminosity, the mass of the central black hole is given by the bulge/BH
mass correlation \citep{mago98,ferr00,hari04}, and that each nuclear source
has a maximum luminosity given by the Eddington limit.  The spread in the
nuclear-to-host luminosity ratio distribution would then be largely produced
by different accretion rates, implying $L/L_{\mathrm{Edd}}$ to range from
$\sim 0.01$ to $\sim 1$ (assuming a fixed radiative efficiency for all the
objects).

Figure \ref{HN2} shows the distribution of the nuclear absolute magnitudes
versus the absolute magnitudes of the spheroidal hosts for our sample (solid
circles), together with different samples of type 1 AGNs obtained from
literature.  We have taken the bulge luminosities whenever the bulge/disk
components of the hosts have been decoupled \citep[e.g.][]{scha00}.  The
\cite{jahn02a} sample has been included for a comparison with another complete
sample of slightly brighter AGNs, at a lower redshift range, $z<0.2$ (open
squares, 19 objects).
We have included two well-defined samples of different families of even
brighter AGNs \citep{dunl03,floy03}, that include radio-loud and radio-quiet
quasars (open stars and open crosses, 19 and 14 objects respectively).
The compilation by \cite{hami02} bundles a large collection of HST 
observations, but it is not, in any sense, a complete or 
well-defined sample (triangles, 70 objects).
Finally we have included an X-ray selected sample of type 1 AGNs, from \cite{scha00}.
Despite the large dispersion, it is clear that all the samples follow the same
trend.

In this figure we have overplotted the expected locations for AGN emitting at
certain fractions of the Eddington limit, assuming a luminosity dependent
bulge mass-to-light ratio, following \cite{dunl03} and \cite{floy03}. Despite
a few AGN that appear to show super-Eddington luminosities, the distribution
of points is in agreement with a maximum of $L/L_{\mathrm{Edd}}\simeq 1$.  The
mean distribution follows mainly the expected location for an object emitting
at $\sim$10~\% of $L_{\mathrm{Edd}}$, (dashed line), while $\sim$90~\% of the
objects are confined within a range of fuelling efficiencies between 100~\%
and 1~\% of the Eddington limit (solid and dashed-dotted line).  There is no
evidence that the upper limit to the nuclear luminosity at given host
luminosity could depend on either the host luminosity or redshift. 

As discussed above, the mass-to-light ratio depends on the $U-V$ color
\citep[e.g.,][]{bell01}. At a given luminosity, blue galaxies are less massive
than red ones. If host galaxy mass is translated into expected black hole
mass, this implies that the Eddington limit for the nucleus of a blue host
galaxy is lower than that for the nucleus of a red host galaxy with the same
luminosity. The Eddington limit plotted in figure \ref{HN2} corresponds to a
red spheroidal, and therefore, is an upper limit. As we quoted in Section 5.1,
the blue colors of the host galaxies are most probably produced by a recent
star formation involving less than a $\sim$30\% of the overall host mass.  In
that case the Eddington limit would be $\sim$0.3 mag fainter for the bluest
host galaxies of our sample, which does not change the overall results.

\section{Summary and conclusions}

We have analyzed a flux-limited sample of 15 intermediate luminosity type 1
AGNs from \combo, at $0.5<z<1.1$ using the \gems\ deep F606W and F850LP-band
ACS images. We decomposed he host and nuclear components and detected the host
in all the F850LP-band images (and in 13 of the F606W-band images). A
morphological analysis has been performed to determine if the hosts were bulge
or disk-dominated objects. $\sim$80~\% of the hosts are spheroidal galaxies,
and only 1 object can be clearly classified as a disk dominated galaxy. We
found evidence for on-going mergers in $\sim$25~\% of the early types and
$\sim$30~\% of the late type hosts.

The host galaxies show a wide range of $\V-\Z$ colors, indicating the
presence of different stellar populations. 
We found that a large fraction of the early-type hosts, $\sim 50-70$\%, 
show distinct blue colors when comparing with the red sequence inactive 
early-type galaxies studied by \gems\ at the same redshift range. 
The color and color-magnitude distributions of the AGN hosts are significantly 
different from those of red sequence galaxies.  
On the other hand, the absolute mangitudes, effective radii and stellar masses 
of these objects are remarkably similar to normal elliptical galaxies. 
Therefore, a significant fraction of the AGN in our sample seem to be 
located in \emph{young spheroids}.

We considered two different simplified scenarios that can explain the observed 
color distributions: (a) a single stellar population with a mean age of close
to 1~Gyr, or (b) a mix of an underlying old stellar population plus a 
small mass fraction in a young component.  Both can describe the observed colors 
well. On the other hand, the early-type host galaxies are structurally similar 
to well-assembled elliptical galaxies, following the luminosity-size relation 
of early-type galaxies. This indicates, most probably, that the process 
that has generated the blue colors has not strongly altered neither the 
luminosity ($\sim$mass) and the effective radius of these galaxies. 
Taking into account this result, it seems that a mixed population, 
dominated by an underlying population of old stars, fits the observations better. 
This underlying population would reflect the population of the parent galaxy.

Our results are in excellent agreement with the results from the study of a
large sample of type 1 and type 2 AGNs from the SDSS at lower redshift
\citep{kauf03}. They noted the existence of a population of \emph{young bulge}
galaxies ($\sim$30\% of their sample), in a state of post-starburst, and a
population of even younger mergers ($\sim$40~\% of their sample). Recent
results by \cite{jahn02a} and \cite{koti04} also found that low-$z$ host
galaxies are bluer than field galaxies. While on the other hand, \cite{dunl03}
and \cite{floy03} did not find evidence for distinctly blue colors in their
sample of high-luminosity low-$z$ AGNs ($z<0.3$), their color range is less
sensitive to young stellar populations at this redshift ($R-K$), especially if
there is an underlying population of old stars.

A possible explanation to the connection between nuclear and star formation
activity in apparently normal elliptical galaxies could be an enrichment of
gas produced by a merging process or the infall and concentration of gas
produced by a minor interaction. Since the fraction of merger galaxies in our
sample seems not to be remarkably larger than in inactive galaxies, the
merger/interaction should have happened in a timescale long enough to enable
the galaxy to relax morphologically before the current AGN activity.

\section*{Acknowledgments}

Support for the GEMS project was provided by NASA through grant number GO-9500
from the Space Telescope Science Institute, which is operated by the
Association of Universities for Research in Astronomy, Inc. for NASA, under
contract NAS5-26555.

EFB and SFS ackowledge financial support provided through
the European Community's Human Potential Program under contract
HPRN-CT-2002-00316, SISCO (EFB) and HPRN-CT-2002-00305, Euro3D RTN (SFS).

CW was supported by a PPARC Advanced Fellowship.

SJ acknowledges support from  the National Aeronautics
and Space Administration (NASA) under LTSA Grant  NAG5-13063
issued through the Office of Space Science.

DHM\ acknowledges support from the National Aeronautics
and Space Administration (NASA) under LTSA Grant NAG5-13102
issued through the Office of Space Science.

The work of CYP was performed in part under contract with the Jet Propulsion
Laboratory (JPL) funded by NASA through the Michelson Fellowship Program.

SFS wants to acknowledges Dr.M.M.Roth for supporting him in this work.

We would like to thanks the anonymous referee for his interesting and useful
remarks that have improved the quality of this article.

\label{lastpage}

\clearpage

\appendix

\section{AGN images and Surface Brightness. Notes on Individual Objects}
\label{a1}

\begin{figure*}
 \includegraphics[width=\hsize]{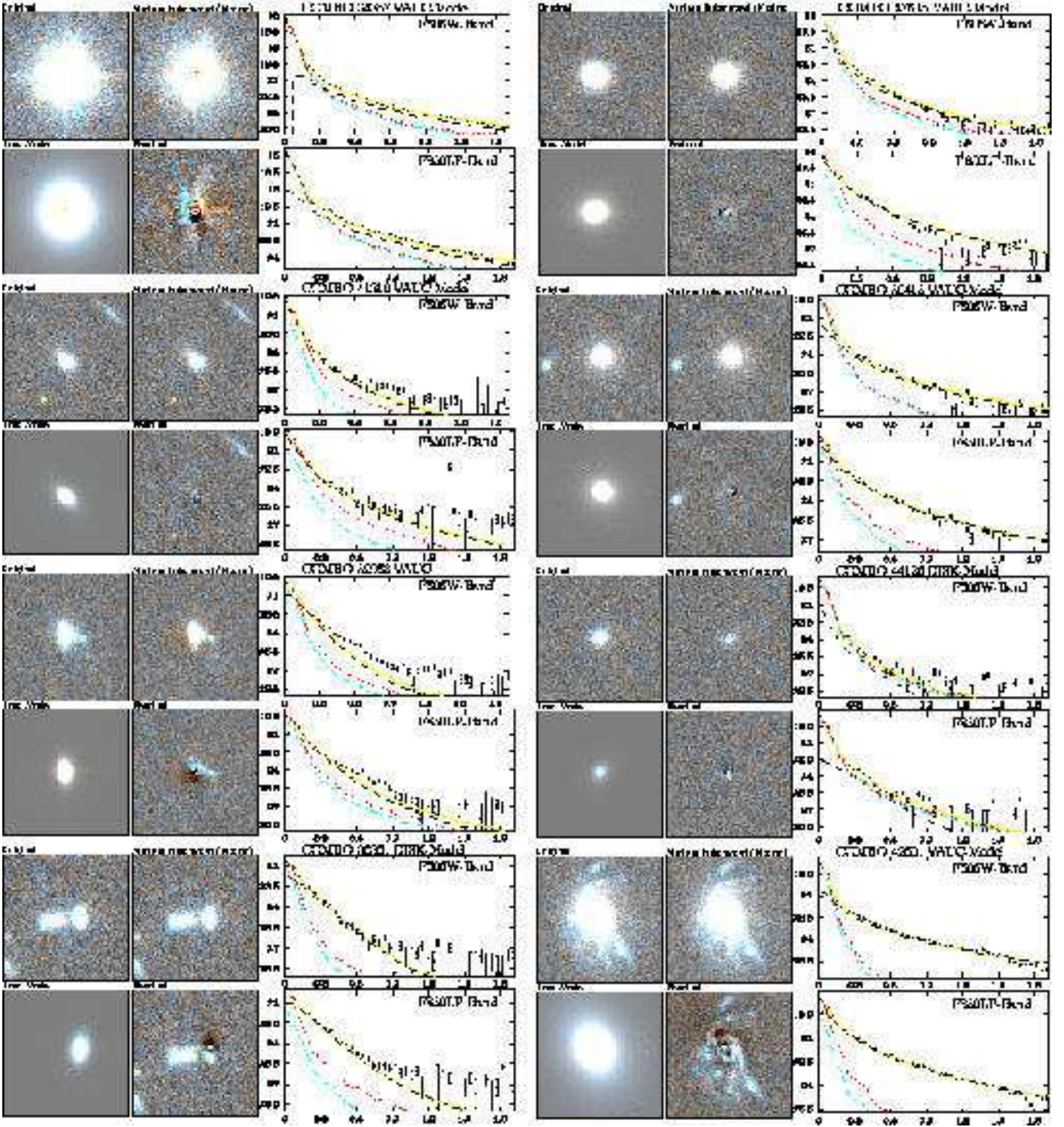}
 \caption{\label{maps} 
   Each of the 15 panels, one per object, shows a color image
   created using the F606W and F850LP-band images for the original postage
   image of the object (top-left), the \emph{restored} host (top-right), the
   host galaxy model (bottom-left), and residual image from the 2D fitting
   (bottom-right).  To the right, we show the surface brightness profile of the
   object (black dots), together with the profile of the best fit (solid line,
   yellow), of the derived nucleus (dotted-dash line, blue), and 
   of the best fitting host model (dashed line, black). For comparison, we have
   included the profiles of the \emph{restored} host galaxy image, together with the
   profiles of the best-fit model and of the peak-scaled PSF substraction (dotted
   line, red).  All surface brightness profiles are in mag~arcsec$^{-2}$, the radii 
   are in arcsec.
   The model fit is marked with a DISK label for an exponential profile, and
   with a VAUC for a de Vaucouleurs profile.
 }
 \end{figure*}

Figure \ref{maps} shows the color images and surface brightness profiles for
all the 15 objects in the sample. We present here detailed notes on individual
objects:

\noindent {\bf  COMBO 34357:} This is the lowest redshift object of our sample. 
Its host is clearly resolved, being a blue spheroidal galaxy, without any clear trace of
major disturbance. The residual image shows a clumpy
structure near the very center that could be traces of an old merging event,
 although it is too close to the central area to be distinguished from a PSF
 subtraction effect.

 \noindent {\bf  COMBO 41310:} A clearly resolved object. The host galaxy is as 
 bright as the nucleus in both
 bands. The galaxy shows a clear spheroidal morphology, without evidence of
 recent interactions. There is a projected companion in the field of view.

 \noindent {\bf  COMBO 52963:} A clearly resolved object. There is evidence for
 a recent or onging merging event, in the form of blue compact clumpy structures
 in both the residual and the original image. The host galaxy is of
 early morphological type.

 \noindent {\bf  COMBO 36361:} A multiple system, comprising three major clumps of blue
 emission, and two nearby companions. There is clear evidence of a recent or
 ongoing merger.  The morphological classification (late-type) is rather
 uncertain, due to the contamination from the nearby companions and
 substructures, some of them as bright as the host galaxy itself.

 \noindent {\bf  COMBO 47615:} A spheroidal host galaxy. There are neither
 appreciable substructures in the residual image not close companions.

 \noindent {\bf  COMBO 50415:} Spheroidal host galaxy. There is a nearby blue and compact
 companion, and maybe blue low surface brightness substructure in the vicinity
 of the host galaxy.

 \noindent {\bf  COMBO 44126:} The most compact and faintest host galaxy in our sample.
 The morphological classification (late-type) is uncertain due to the low
 brightness and size.

 \noindent {\bf  COMBO 42601:} One of the biggest 
 and brightest hosts of our sample. It is clearly classified as an
 early type galaxy, dominated by a large and blue spheriodal component. The
 residual image shows bright clumps and filaments, most probably due to recent
 star formation.

 \noindent {\bf  COMBO 48284:} Big and bright host galaxy, 
 clearly resolved. It shows a clearly spheroidal morphology, without traces of
 disturbance.

 \noindent {\bf  COMBO 39432:} A compact spheroidal host galaxy.
 No irregularities in its morphology.

 \noindent {\bf  COMBO 31898:} A compact spheroidal host galaxy.
 This is the reddest object of our sample, which can be appreciated
 in the color images.

 \noindent {\bf  COMBO 15731:} One of the most beautiful objects of our sample. The host
 galaxy seems to be dominated by a central bulge, although both the residual
 and the host {\it restored} image show evidence of what could be two spiral
 arms.  We have classified this object as an early-type galaxy, since it is
 clearly bulge dominated.  It presents a nearby blue and compact companion,
 which may be interacting with the host. A possible bridge between the host and
 the companion is seen. It is difficult to clarify if the apparent arms are
 part of a real disk or structures produced by a recent collision.

 \noindent {\bf  COMBO 50997:} The host galaxy of this object has been only detected in
 the F850LP-band.  It is a compact spheroidal galaxy. It shows a close
 projected companion as big and bright as the host galaxy.

 \noindent {\bf  COMBO 49298:} A clear late-type galaxy. 
 The spiral arms are clearly seen in the all the images.

 \noindent {\bf  COMBO 43151:} This object is the highest redshift 
 object of our sample.
 Its host galaxy has been detected only in the F850LP band, being a compact
 spheroidal galaxy, without evidence for disturbances.

 \addtocounter{figure}{-1}

 \begin{figure*}
 \includegraphics[width=\hsize]{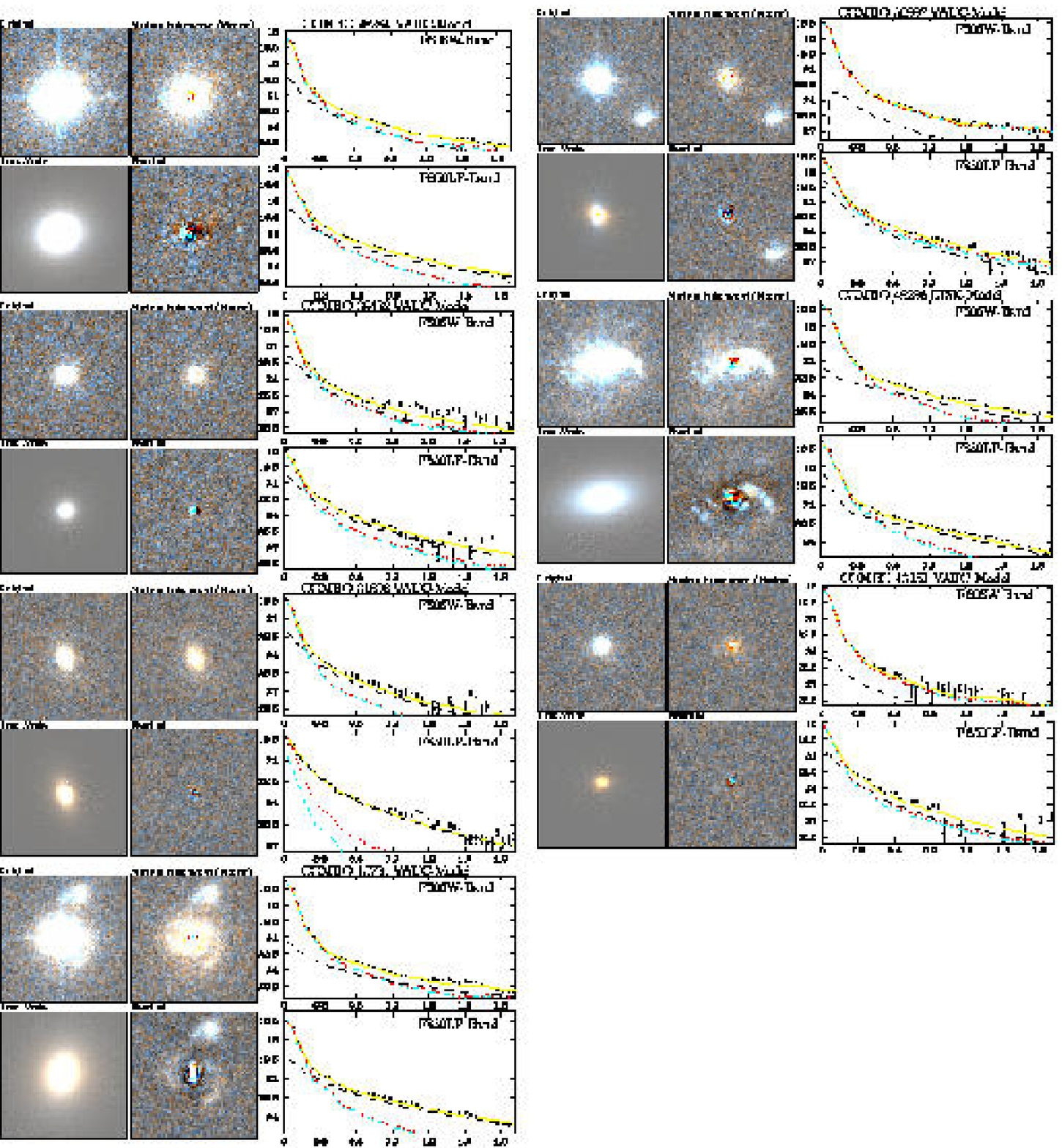}
 \caption{\label{maps1} 
 Continued
 }
 \end{figure*}

 \newpage

\section{From observed to rest-frame: empirical $k$-corrections}
\label{a2}

\begin{figure*}
\includegraphics[angle=-90,width=\hsize]{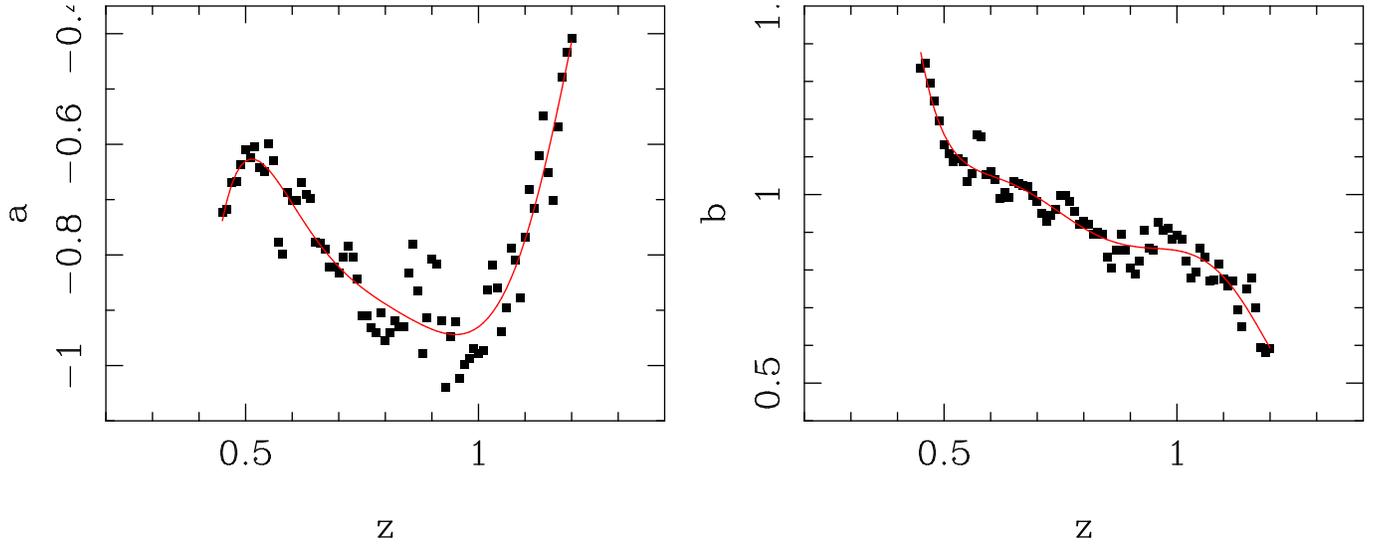}
\caption{\label{k_cor_0e} 
  Distributions of the average $a$ and $b$ parameters of the linear
  regressions between the observed F606W-F850LP and the rest-frame $U-V$
  colors as a function of the redshifts, for the $\sim$4000 inactive galaxies
  in the \gems\ field with \combo\ photometry.  The solid lines show 
  high-order polynomial fits to these distributions, the parameters
  of which can be found in Table~\ref{tab_grid_colors}.
}
\end{figure*}

\begin{figure*}
\includegraphics[angle=-90,width=\hsize]{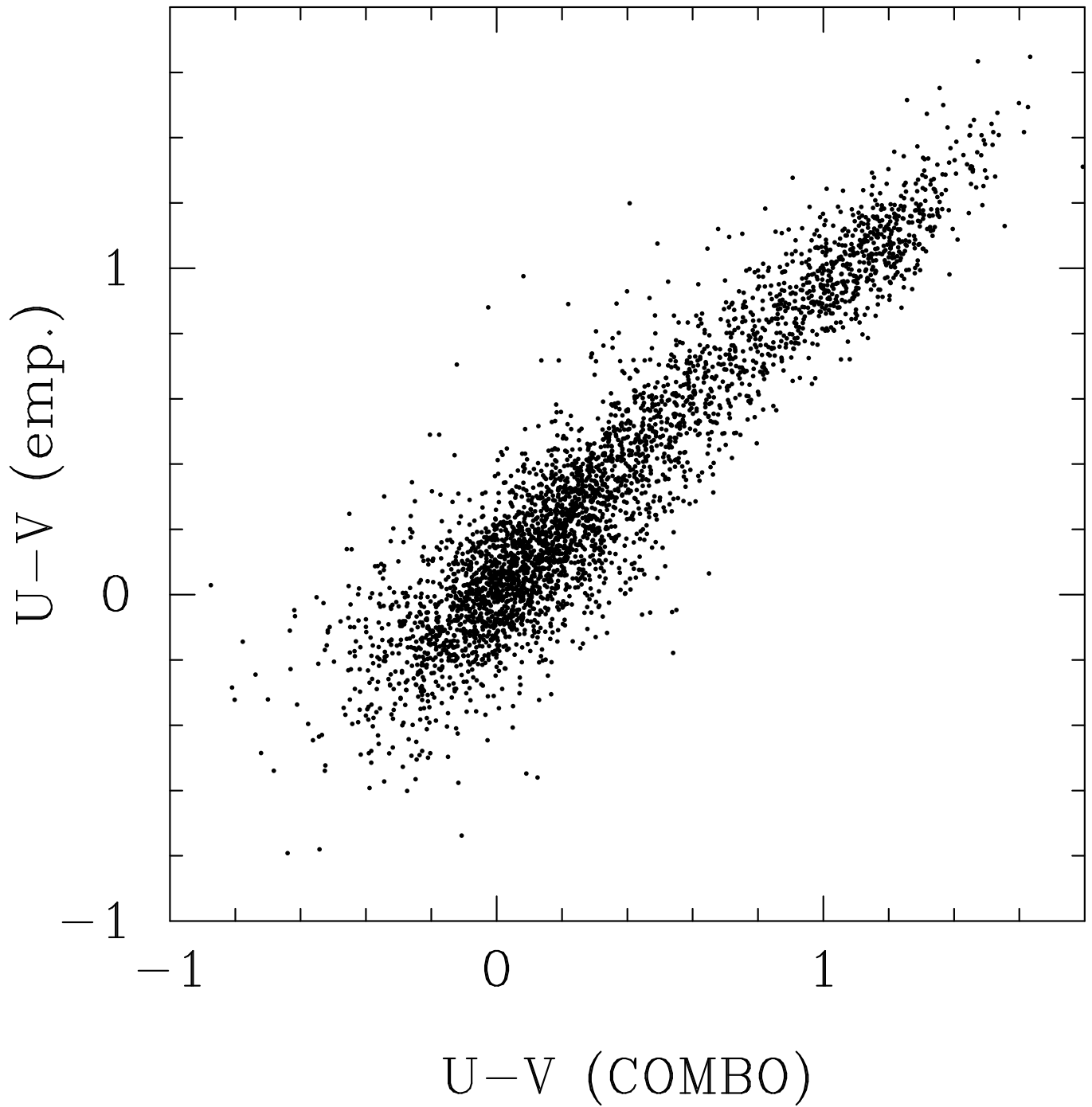}
\caption{\label{k_cor_3} 
  Comparison between rest-frame $U-V$ colors of inactive galaxies from
  \combo\ multiband photometry and the $U-V$ values estimated from
  observed \gems\ F606W-F850LP, using the transformation described in the text.
}
\end{figure*}

We calculated $k$-corrections for the host galaxies by comparing them with
galaxies of the same magnitude, redshift and colors from the \combo\ survey
where spectral energy distributions (SEDs) are available for all objects.
First, we derived the rest-frame $U-V$ colors from the observed $\V-\Z$. For
doing so we performed aperture photometry on the $\sim$4000 inactive galaxies
in the \gems\ field at the same redshift as our objects. We split this sample
of galaxies in different redshift slices, each with a width of 0.02 and
stepped by 0.01 in redshift.  For each slice we found a good linear
correlation between the observed $\V-\Z$ and the rest-frame $U-V$ colors. We
computed linear regression solutions for the transformation
\begin{equation}
  U-V = a(z) + b(z)\times(\V-\Z)
\end{equation}
for each slice separately, considering both forward and backward
transformations and averaging the two.  The rms dispersions were between
$0.12-0.22$ mag. This provided a grid of transformation coefficients $a(z)$
and $b(z)$ as shown in Fig.~\ref{k_cor_0e}.  We then fitted the $a(z)$ and
$b(z)$ relations with high-order polynomials.  Table \ref{tab_grid_colors}
provides the polynomial coefficients needed to reconstruct the $a$ and $b$
parameters and perform the transformation at any redshift.
In order to cross-validate this set of
color transformations, we compared the rest-frame colors derived in this way
with the original \combo\ colors, for the input sample of inactive galaxies
(Figure \ref{k_cor_3}).  We found no systematic differences between both
colors, and a spread of $\sim$0.16 mag (rms).

\begin{figure*}
\includegraphics[angle=-90,width=\hsize]{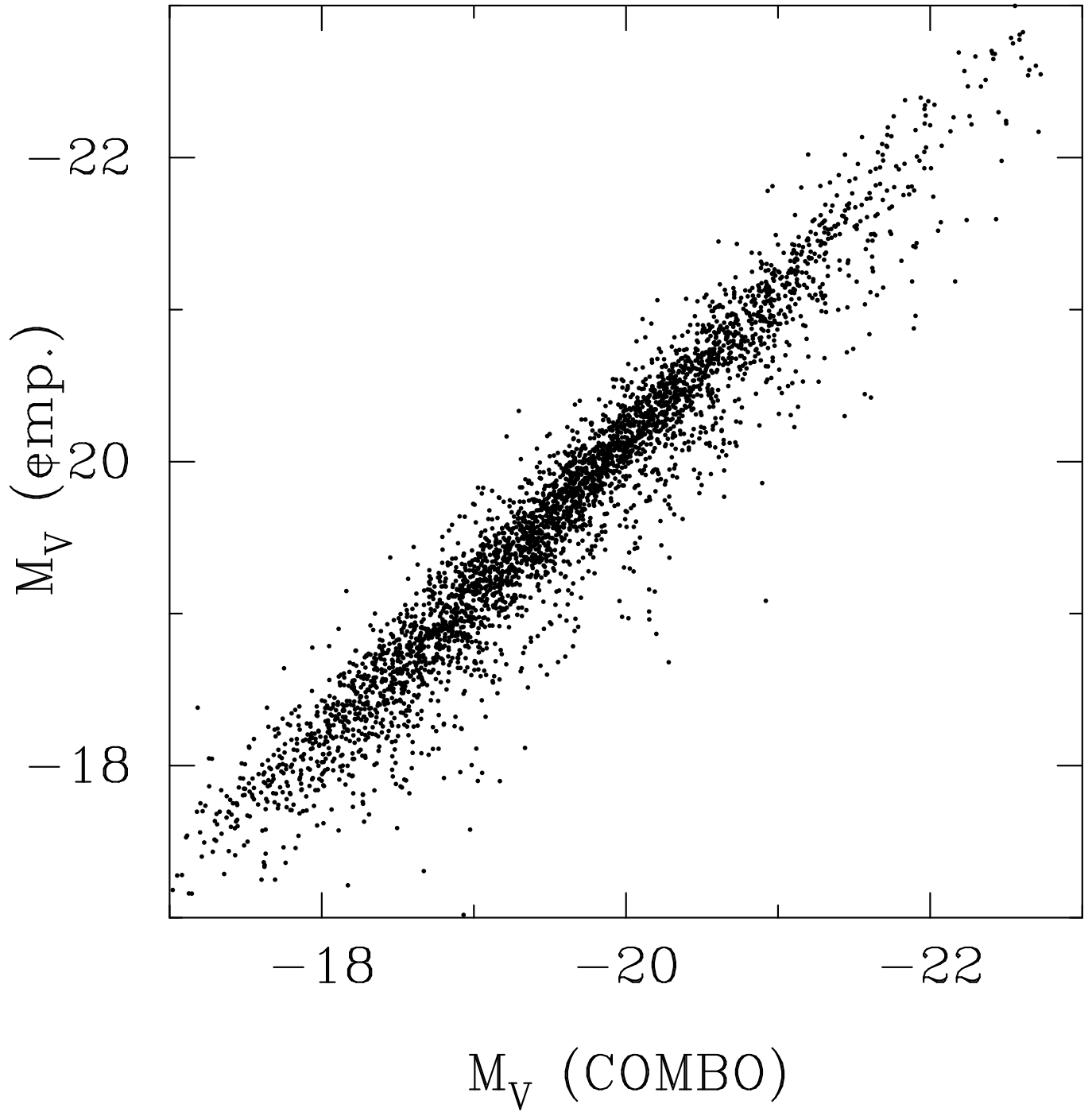}
\caption{\label{k_cor_2} 
  Comparison between absolute magnitudes $M_V$ of inactive galaxies from
  \combo\ multiband photometry and color-dependent $M_V$ estimated from
  observed \gems\ F850LP, using the transformation described in the text.
}
\end{figure*}

We then created a grid of \combo\ $U-V$ colors and redshifts, with a
separation of 0.2 mag in colors and 0.02 in redshifts. Each grid box contained
all objects within a color range of 0.6~mag and a redshift range of 0.1
around the central values of the box. For each grid box we found again
a linear relation between the observed F850LP magnitude and the \combo -based
absolute $V$ band magnitude 
\begin{equation}
  M_V = A(z,U-V) + B(z,U-V)\times \Z
\end{equation}
where the regression coefficients however depend not only on redshift but 
also on the intrinsic color. Note that by using F850LP magnitudes at the 
mean redshift of our objects, we minimize this color term as we 
always sample the SED no far from the rest-frame $V$ band.
The rms dispersions of the $M_V$ values around the linear regression fits are
around 0.15--0.22~mag. We fitted the regression parameters with 2nd order
polynomials as a function of $z$, separately for each color range.  Table
\ref{tab_grid_mag} provides the coefficients required to derive the $A$ and
$B$ regression parameters. Again, we cross-validated this set of
transformations by applying it to the \combo\ inactive galaxies, finding no
systematic differences between the two ways to estimate $M_V$, and a spread of
$\sim$0.28 mag (Figure \ref{k_cor_2}).

\begin{table}
\begin{center}
\caption{Polynomial coefficients describing the redshift dependence
  of color transformation parameters $a=p_n (z)$ and $b=q_n (z)$.}
\label{tab_grid_colors}
\begin{tabular}{rrr}
\tableline\tableline
Order $n$ & $p_n$ & $q_n$  \\
\tableline
0     &    $-$64.0375   &     71.9392\\
1     &    488.2067   &   $-$538.3892\\
2     &  $-$1529.8569   &   1676.0263\\
3     &   2503.5583   &  $-$2730.2606\\
4     &  $-$2262.2580   &   2450.3622\\
5     &   1070.2635   &  $-$1148.4126\\
6     &   $-$206.8062   &    219.5862\\
\tableline
\end{tabular}
\end{center}
\end{table}

\begin{table*}
\begin{center}
\caption{Polynomial coefficients describing the redshift dependence
  of observed to absolute magnitude transformation parameters 
  $A = P_n (z)$ and $B=Q_n(z)$, separately for different rest-frame colors.}
\label{tab_grid_mag}
\begin{tabular}{rrrr|rrrr}
\tableline\tableline
$U-V$&Order $n$ & $P_n$ &  $Q_n$ & $U-V$ & Order $n$ & $P_n$ & $Q_n$ \\
\tableline
$-$0.4    &0      & $-$36.3631    &   0.8845&  0.6    &0      &$-$37.8007     &   0.9602\\
$-$0.4    &1      & $-$26.1132    &   0.9650&  0.6    &1      & $-$2.5629     &  $-$0.1072\\
$-$0.4    &2      &  20.0688    &  $-$0.8527&  0.6    &2      & $-$2.5713     &   0.1569\\
\tableline
$-$0.2    &0      & $-$38.9592    &   1.0106&  0.8    &0      &$-$28.9003     &   0.5637\\
$-$0.2    &1      & $-$11.0029    &   0.2691&  0.8    &1      &$-$28.2281     &   1.0473\\
$-$0.2    &2      &   7.7105    &  $-$0.2934&  0.8    &2      & 13.3033     &  $-$0.5641\\
\tableline
 0.0    &0      &$-$34.9194     &   0.8240&  1.0    &0      &$-$38.8294     &   1.0262\\
 0.0    &1      &$-$16.9380     &   0.5508&  1.0    &1      &  4.4406     &  $-$0.4684\\
 0.0    &2      &  9.3486     &  $-$0.3781&  1.0    &2      &$-$12.5859     &   0.6327\\
\tableline
 0.2    &0      &$-$34.0425     &   0.7915&  1.2    &0      &$-$44.8231     &   1.3100\\
 0.2    &1      &$-$17.9921     &   0.5816&  1.2    &1      & 20.8241     &  $-$1.2308\\
 0.2    &2      &  9.5968     &  $-$0.3823&  1.2    &2      &$-$22.9941     &   1.1100\\
\tableline
 0.4    &0      &$-$33.2426     &   0.7501&  1.4    &0      &$-$57.9184     &   1.9331\\
 0.4    &1      &$-$16.3457     &   0.5202&  1.4    &1      & 55.7256     &  $-$2.8866\\
 0.4    &2      &  6.5599     &  $-$0.2537&  1.4    &2      &$-$44.3401     &   2.1188\\
\tableline
\end{tabular}
\end{center}
\end{table*}

\end{document}